\documentclass[%
reprint,
 amsmath,amssymb,
 aps,
]{revtex4-1}

\usepackage{graphicx}
\usepackage{dcolumn}
\usepackage{bm}

\begin{document}


\title{Observational Constraints of Stellar Collapse:\\Diagnostic Probes of Nature's Extreme Matter Experiment}

\author{Chris L. Fryer}
 \altaffiliation[Also at ]{Physics and Astronomy Department, University of New Mexico, Albuquerque, NM 87545; Department of Physics, The University of Arizona, Tucson, AZ, 85721}
 \email{fryer@lanl.gov}
\author{Wesley Even}%
\affiliation{CCS Division, Los Alamos National Laboratory, Los Alamos, NM 87545, USA}%

\author{Brian W. Grefenstette}
\affiliation{Space Radiation Lab, California Institute of Technology, Pasadena, CA 91125, USA}%
\author{Tsing-Wai Wong}
\affiliation{Center for Interdisciplinary Exploration and Research in Astrophysics (CIERA) \& Department of Physics and Astronomy, Northwestern University,
2145 Sheridan Road, Evanston, IL 60208}%
\affiliation{Harvard-Smithsonian Center for Astrophysics, 60 Garden St., Cambridge, MA 02138, USA}%

\date{\today}

\begin{abstract}
Supernovae are Nature's high-energy, high density laboratory
experiments, reaching densities in excess of nuclear densities and
temperatures above 10\,MeV.  Astronomers have built up a suite of
diagnostics to study these supernovae.  If we can utilize these
diagnostics, and tie them together with a theoretical understanding of
supernova physics, we can use these cosmic explosions to study the
nature of matter at these extreme densities and temperatures.
Capitalizing on these diagnostics will require understanding a wide
range of additional physics.  Here we review the diagnostics and the
physics neeeded to use them to learn about the supernova engine, and
ultimate nuclear physics.

\end{abstract}

\pacs{26.30,97.60}
\maketitle


\section{\label{sec:introduction}Introduction:  Supernovae as Extreme-Matter Experiments}

Core-Collapse Supernovae are among the most powerful explosions in the
universe, produced when the core of a massive star is no longer able
to support itself and collapses.  The collapse continues until the
core reaches nuclear densities at which point nuclear forces and
neutron degeneracy pressure halt the collapse.  The compression
produces densities in excess of nuclear densities and temperatures
above 10\,MeV.  Although laboratory experiments such as PREx
\cite{abrahamyan12} might allow scientists to probe properties of the
neutron, core-collapse supernova have the potential to probe the broader 
properties of nuclear matter at densities above nuclear.

Nuclear physics is at the heart of the core-collapse supernova
problem.  The behavior of matter at nuclear densities determines the
depth of the bounce which, in turn, can alter the strength of the
bounce shock causing a prompt explosion\cite{baron87}.  Although this
prompt explosion mechanism is no longer favored, the behavior of
matter at nuclear densities still plays a role in the explosion,
defining the neutrino luminosity and spectra, determining the maximum
compression of the core and altering the extent of the bounce shock.  
Similarly, uncertainties in the nuclear reaction rates alter both the 
explosive and r-process yields in supernovae.

In this paper, we review core-collapse supernovae from the
point-of-view of a nuclear physics experiment.  As with many
laboratory experiments, nuclear physics is only one piece of the
core-collapse supernova problem (see review by Hix et al. in this
issue), and all this physics must be understood and/or its
uncertainties limited if we are to study any piece of the physics.
Fortunately, there are a broad set of diagnostics probing
core-collapse supernovae ranging from direct probes of the core
limited to only the most nearby supernovae (neutrinos and
gravitational waves) to integrated probes that include many objects
(nucleosynthetic yields or the mass distribution of compact remnants).
But to truly take advantage of this data, theoretical models are
critical.  New theory models of the progenitors, the supernova engine
and its subsequent explosions have been advancing the physics and
strengthening our ability to use these diagnostics to probe the
physics behind core-collapse supernovae.  This approach is not so
different than many laboratory experiments.  Although experimentalists
strive to design an experiment to directly observe the intended
physics, most experiments are driven by a wide range of physics and
theoretical models are often needed to reduce the uncertainties and
interpret the results.

Before we discuss the diagnostics in detail, let's review the basic
supernova engine mechanism and the stages at which the diagnostics
help us pin down this engine.  We will revisit aspects of this
explosion mechanism throughout the paper as they pertain to specific
diagnostic probes.  The supernova progenitor, a star more massive than
$\sim 8M_\odot$ progresses through a series of burning phases,
building up a dense core. For most pre-supernova progenitors, an iron
core is produced that continues to grow until it can no longer support
itself, collapsing to form a neutron star. Some super-asymptotic giant
branch stars can collapse without forming an iron core, the so-called
electron capture supernovae\cite{herwig05}. Supernova progenitors can
be observed by searching for the absence of progenitors in survey data
after the supernova fades and the list of these observed stars has
increased steadily over the past decade (see
Sec.~\ref{sec:progenitor}).

The iron core is supported by thermal and electron degeneracy
pressure. As the mass increases, the core contracts until electron
capture and the dissociation of iron atoms remove both these pressure
sources, causing the core to collapse. The compression accelerates the
capture and dissociation, leading quickly to a runaway collapse. The
collapse continues until the core reaches nuclear densities where
neutron degeneracy pressure and nuclear forces halt the collapse,
causing the imploding star to bounce. The bounce shock stalls and the
current "standard" picture of supernovae argues that convective
instabilities (e.g. Rayleigh-Taylor\cite{herant94}, advective-vortical
instabilities\cite{blondin03}, or advective-acoustic\cite{foglizzo07})
revive the shock, driving an explosion.

Neutrinos emitted in the core increase during the collapse but, in the
extreme densities at bounce can be trapped in the core.  As the bounce
shock expands, the neutrinos leak out, producing the strong burst of
neutrinos and a near-direct probe of the behavior of matter at nuclear
densities.  The neutrino signal persists as the neutron star cools and
can be further enhanced by late-time accretion (caused as material
initially swept up in the shock is unable to escape the gravitational
pull of the neutron star) onto the supernova.  Gravitational waves
also probe this core region during the bounce and convective engine
time frame, but are more sensitive mass motions caused by rotation and
convection.  Elements are synthesized in a variety of processes:
explosive nucleosynthesis as the shock moves through the silicon layer
in the star causing fusion up through iron peak elements, r-process
elements produced in winds from the cooling proto-neutron star or from
secondary ejecta caused by accretion onto the newly formed neutron
star.  Measuring these nucleosynthetic yields in stars or asteroids
provides another clue into the supernova engine (albeit in an
integrated sense).

As the supernova shock breaks out of the star, photons trapped in the
shock are also able to leak out.  This shock breakout marks the first
optical emission from the supernova explosion.  The shock continues to
expand, producing the bright emission whose details characterize the
broad range of Ib,Ic, IIp, IIn, IIl, IIb observed supernova classes.
Observations of the breakout probes the photosphere of the star,
whereas the peak and late-time light curve probe both characteristics
of the star (photosphere, circumstellar medium, sometimes mass) as
well as the yield of $^{56}$Ni.  

Supernova remnants, both the compact collapsed cores (neutron stars and
black holes) and the ejecta remnants provide further diagnostics of
the supernova explosion.  Hundreds of years after the explosion, the
supernova ejecta is still visible as a remnant.  The supernova
ejecta-remnants provide yet further clues to the explosion asymmetry
and nucleosynthetic yields.  The mass distribution and spin of compact
remnants (the black hole or neutron star formed during collapse) can
also provide clues in understanding the supernova engine.  In this
review, we study all of these diagnostics.  First, we focus on the
most direct diagnostics of the nuclear physics: neutrinos
(Sec.\ref{sec:neutrinos}) that probe the collapsed core with the
neutrinosphere touching the surface of the newly-formed neutron star
and gravitational waves (Sec.\ref{sec:GW}) that probe the matter
motion near the proto-neutron star.  We then turn to the more
integrated diagnostics.  These diagnostics are much more common, but
analyzing the data requires much more complex models with integrated
physics.  To constrain these diagnostics, we must understand the
progenitors (Sec.~\ref{sec:progenitor}) and obtain as much data as
possible on the supernova outburst itself including light curves and
spectra (Sec.~\ref{sec:SNLC}), the supernova ejecta
(Sec.~\ref{sec:ejectaremnant}), mass distributions of the compact
remnants (Sec.~\ref{sec:compactremnant}), and the integrated
nucleosynthetic yield estimates (Sec.~\ref{sec:nucobs}).  As an
example of the strength of these combined diagnostics, we review the
study of r-process elements from core-collapse supernovae
(Sec.~\ref{sec:postexp}).  

\section{Neutrinos}
\label{sec:neutrinos}

In an ideal experiment, we develop direct probes of the physics of
interest (in our case, the behavior of matter at nuclear densities).
If no additional physics alters our probe (or if we understand that
additional physics perfectly), we can measure the experiment's goal
physics directly.  For our nuclear-physics supernova experiment,
neutrinos are that direct diagnostic.  The neutrinosphere, the surface
of last scattering for neutrinos, occurs just above the edge of the
proto-neutron star formed in the collapse of a massive star.  As such,
it provides a detailed probe of the conditions of the core and the
behavior of matter at nuclear densities.  In addition, understanding
neutrino interactions themselves is an important open question in
nuclear and particle physics.

A broad set of neutrino detectors sensitive to the $\sim$\,10MeV
neutrinos characteristic of core-collapse supernovae are either
operating or under development.  These detectors take advantage of a
broad set of technologies: liquid scintillation, water Cherenkov, and
liquid Argon detectors.  A review of the current state of our
understanding of neutrinos and current and upcoming neutrino detectors
has recently been completed\cite{degouvea13}.  Here we will briefly
summarize this review and then focus on how observations of supernova
neutrinos probe nuclear and neutrino physics.

For liquid scintilation detectors, the Borexino\cite{alimonti09} and
KamLAND\cite{mitsui11} detectors continue to operate, SNO$+$\cite{kraus10} and
JUNO\cite{balantekin13} experiments are under construction, and the
Hanohano\cite{maricic11} and LENA\cite{wurm12} experments are in the proposal phase.
Super-Kamiokande\cite{abe11} and SNO\cite{aharmim13} are the leading water Cherenkov
detectors and the Hyper-K (the upgrade to Super-Kamiokande) detector
is under research and development.  The shear size of Hyper-K will
yield 250,000 interactions from a Galactic supernova, providing a
detailed signal ($\sim$ 250,000 interactions) to study many nuances of
neutrino and nuclear physics (if we can use diagnostics to constrain
the progenitor, rotation and aspects of the explosion engine).  The
Large Baseline Neutrino Experiment is pursuing liquid argon detector
technology that will be sensitive to $\nu_e$ interactions, probing 
the physics signatures predicted for these neutrinos.

Supernova probes of neutrino physics include many-body interactions,
nuclear correlation effects, mean-field approximations
(e.g.\cite{roberts12a,pinedo12}).  Detections of the neutrinos,
especially if we can place constraints on both the neutrino and
anti-neutrino spectra, can constrain this physics.  These neutrino
effects play a role not only in the neutrino heating, but also the
nucleosynthetic yields.  Coupling neutrino observations with detailed
yield analyses (see Sec.\ref{sec:postexp}) will place strong
constraints on this physics.  Many of these effects can also be
studied after the launch of the supernova in the cooling neutron star
(e.g.\cite{roberts12b,reddy99}).  Neutrinos from material falling back
on the newly formed neutron star can alter the late-time neutrino
signal from cooling neutron stars, limiting what we can learn from the
neutrino signal\cite{fryer09}.  Fortunately, we can place constraints
on this fallback by using secondary diagnstics.  For example, light
curve observations constrain both the explosion energy and $^{56}$Ni
yield.  Coupled with theoretical models of the explosion, we can use
the light-curve observations to place constraints on the total
fallback.  The theory models of fallback can, in turn, be tested by
the mass distribution of compact remnants (a third diagnostic).

\begin{figure}
\includegraphics[width=8cm]{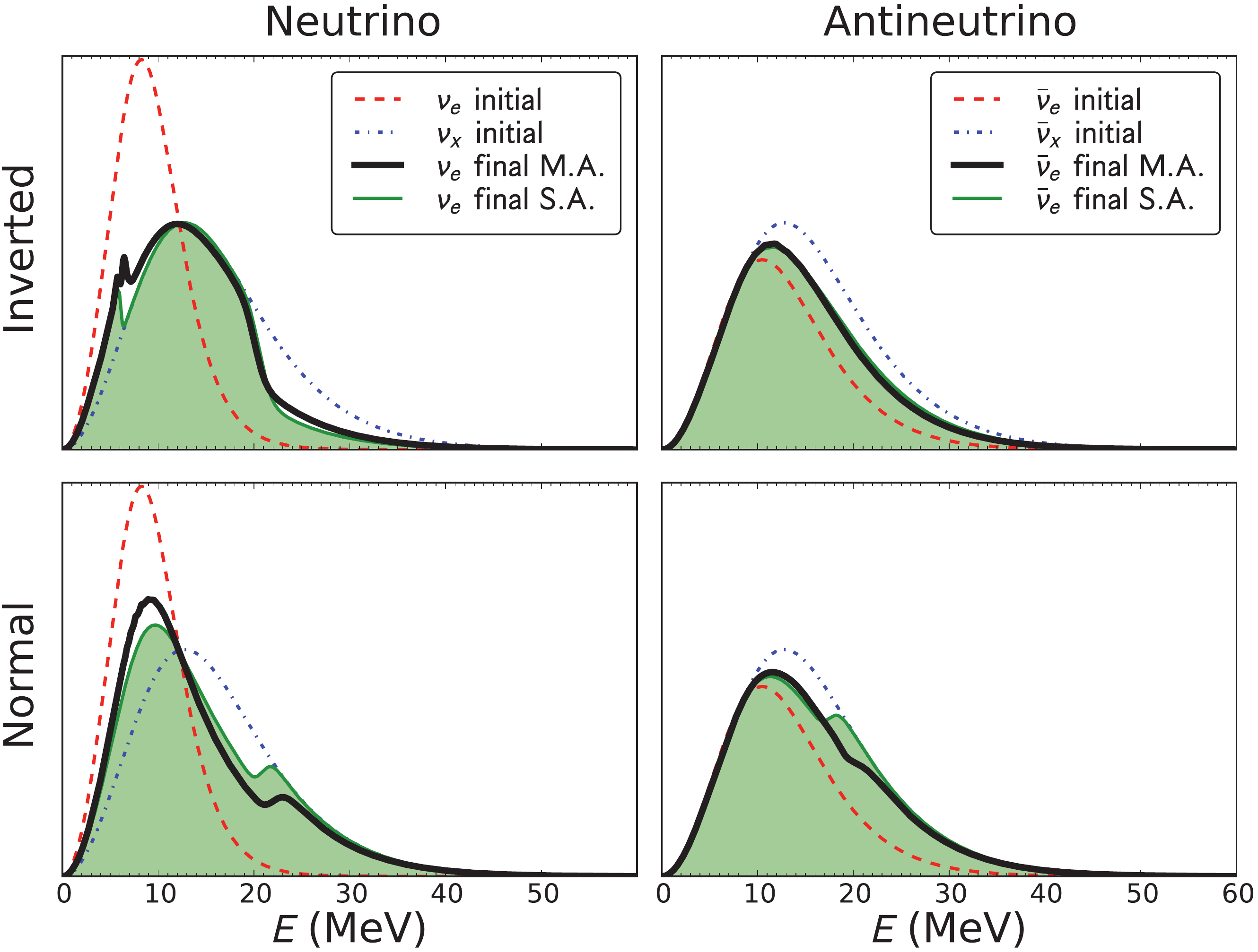}
\caption{Spectra of $\nu_{e}$ \emph{(left)} and $\bar\nu_{e}$
  (\emph{right}) at infinity, for both IH \emph{(top)} and NH
  \emph{(bottom)}. Multiangle (M.A.) results are shown with thick
  solid curves, single-angle (S.A.) with filled regions. Initial
  spectra are also shown, as marked.  Reprinted with permission from
  H. Duan \& A. Friedland, Physical Review Letters, Vol 106, 091101
  (2011)\cite{duan11a}}
\label{fig:neutoscill}
\end{figure}

The detection of neutrino oscillations and a non-zero neutrino mass
opens up a number of possibilities in neutrino physics.  A major focus
of the last 5 years has been the study the flavor-state coupling of
neutrinos in the core.  The high number densities in the core allow
mutual coherent scattering, resulting in many-body flavor
evolution\cite{duan06,fogli07,raffelt07a,raffelt07b,esteban08,duan09,dasgupta09,duan10,duan11a}.
Figure~\ref{fig:neutoscill} shows an example of how the observed neutrino spectra can
change in core-collapse supernovae\cite{duan11a}.  If we are able to
measure a detailed supernova spectrum, we will place strong
constraints on this oscillation physics.  In addition, these neutrino
oscillations will alter the nucleosynthetic yields and, if we can
constrain the astrophysical uncertainties, we may be able to use these
yields to constrain neutrino physics\cite{duan11b}.

The neutrino spectra can also be affected by oscillations in the
turbulence behind the outgoing supernova shock and signatures of the
Mikheyev-Smirnov-Wolfenstein (MSW) matter effect can be observed in
the neutrino spectra (for a recent discussion, see\cite{borriello13}.
These oscillations will alter the neutrino spectra and a detailed
spectra can be used to better understand the turbulence and physics
models for these oscillations.  If we can constrain the turbulence
physics through other observations (e.g. ejecta remnants -
section~\ref{sec:ejectaremnant}), the neutrino spectrum places a
strong constraint on the MSW effect.

Although we will be able to achieve the detailed spectra needed to
test theories of neutrino oscillation with a Galactic supernova, what
we can learn from supernovae beyond the Milky Way is limited.
Hyper-Kamiokande will detect a hand-full of neutrinos, perhaps even
out to the Virgo cluster, but probing nuclear or neutrino physics from
this sparse signal will be difficult.  Even with Galactic supernovae,
we must be able to extract the physics from a variety of physics
beyond the nuclear and neutrino physics (e.g. hydrodynamics, initial
condition uncertainties, etc.).  In this section, we discussed three
instances where the neutrino signal can tell us more about the physics
by including additional diagnostics: using additional diagnostics to
constrain the error from fallback accretion, using nucleosynthetic
yields to provide early insight into neutrino oscillations, and using
ejecta remnants to constrain turbulence in the explosion to better
analyze the MSW effect.  In this manner, a full suite of diagnostics
is crucial to learning physics from supernovae.  In the next sections,
we review these additional diagnostics.

\section{Gravitational Waves}
\label{sec:GW}

Gravitational waves, like neutrinos, are also nearly direct probes of the
central engine behind supernovae.  Whereas neutrinos probe the
temperature/density conditions, gravitational waves depend more
strongly on the matter motion and rotation of the core.  Together,
these probes provide broad information on the supernova central
engine.

Ground-based interferometric gravitational-wave detectors are
ideally-suited for the detection of gravitational waves from stellar
collapse.  With proof-of-concept designs demonstrated in the Laser
Interferometric Gravitational-wave Observatory (LIGO) and Virgo
detectors, the next generation of gravitational-wave detectors are
being developed: advanced LIGO\cite{price13}, advanced
Virgo\cite{price13}, KAGRA\cite{aso13}.  The improved sensitivity of
these detectors coupled with their broader geographic locations will
allow gravitational wave detectors in the next decade to observed
detailed signals from Galactic supernovae and pinpoint the locations 
of these supernovae (important if the supernova is obscured).

Gravitational waves are produced when matter is accelerated in a
manner to produce perturbations ($h_{jk}$) in a background spacetime
($g^b_{jk}$).  Typically, scientists estimate the gravitational wave
signal using a multipole expansion, assuming the lowest (quadrupole) 
order piece dominates the signal:
\begin{equation}
h^{\rm TT}_{jk} = \left[ \frac{2}{d} \frac{G}{c^4} \frac{d^2}{dt^2}
  I_{jk}(t-r) + \frac{8}{3d} \frac{G}{c^5}
  \epsilon_{pq(j}\frac{d^2}{dt^2}S_{k)p}(t-r)n_q \right]^TT,
\end{equation}
where $I_{jk}$ and $S_{jk}$ are the mass and current quadrupole
moments of the source, $d$ is the distance from the source to the
point of measurement, $G$ is the gravitational constant, $c$ is the
speed of light, $\frac{d^2}{dt^2}$ is the second second derivative
with respect to time ($t$), $\epsilon_{ijk}$ is the antisymmetric
tensor, and $n_q$ is the unit vector pointing in the propagation
direction.  Parantheses in the subscripts indicate symmetrization over
the enclosed indices and the superscript $TT$ indicates that one is to
take the travsverse-traceless projection.  Note that a signal only
occurs of the acceleration of either the mass or current quadrupole
moment is non-zero.  

A strong a signal requires not only the rapid acceleration of a large
amount of mass, but the rapid acceleration of the quadrupole moment of
that mass.  The large masses and accelerations in core-collapse make
these supernovae ideal sources of gravitational wave emission.
Several reviews exist covering the gravitational wave signals from
core-collapse supernovae, e.g.\cite{fryernew11}, and we will
summarize this literature, focusing on what we can learn from
gravitational waves and how this diagnostic helps us better understand
supernovae and the physics behind supernovae.  Gravitational waves in
core-collapse can be separated into 4 separate phases: the bounce of
the core, the convection above the proto-neutron star (that ultimately
drives the explosion, neutrino emission and convective instabilities
in the cooling neutron star, and the further collapse to a black hole 
in those systems whose core masses exceed the maximum neutron star mass.

\begin{figure}
\includegraphics[width=8cm]{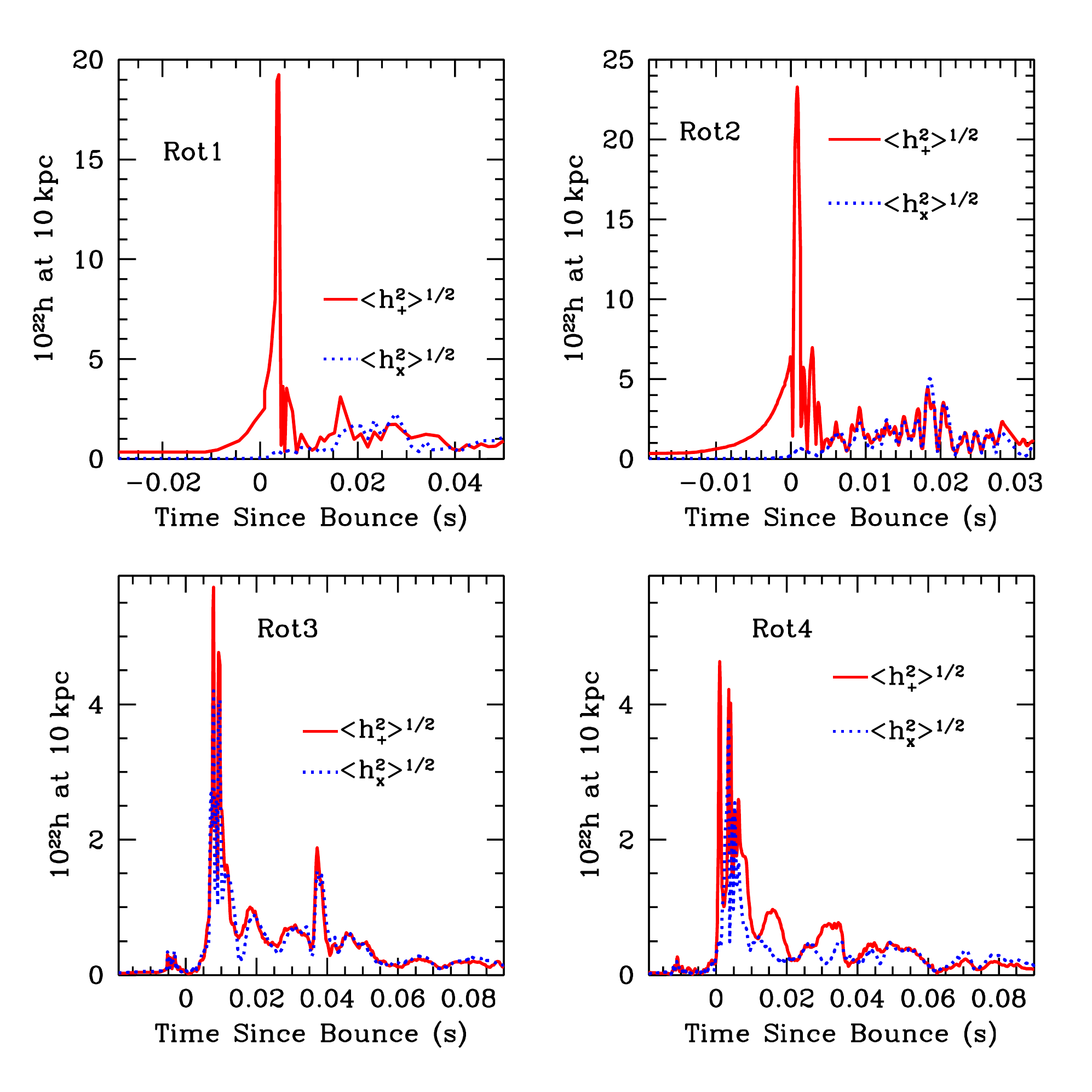}
\caption{The angle averaged wave amplitudes ($<h^2_{+}>^{1/2}$: solid
  line,$<h^2_{x}>^{1/2}$: dotted line) for the mass motions from 4
  rotating supernova models.  Rot2 is the fastest rotating model.
  Rot3 is what would be predicted for a magnetically braked core.
  Rot4 shows the signal from the same star as Rot2, but where the
  rotation was set to zero just before collapse\cite{FHH04}.  Note
  that the gravitational wave signal is a factor of 5 higher in the
  rapidly versus slowly rotating models.  A fast-rotating supernova in
  the Galaxy should be detectable by advanced LIGO.}
\label{fig:gwrot}
\end{figure}

For most simulations, the strongest gravitational wave signals are
produced during the bounce phase of a rotating progenitor star.  In
extremely rapid rotating cores, the proto-neutron star can form bar
mode instabilities or fragment, generating strong, distinct
signatures.  Albeit unlikely, these are among the strongest 
core-collapse signals\cite{FHH02}.

\begin{figure}
\includegraphics[width=8cm]{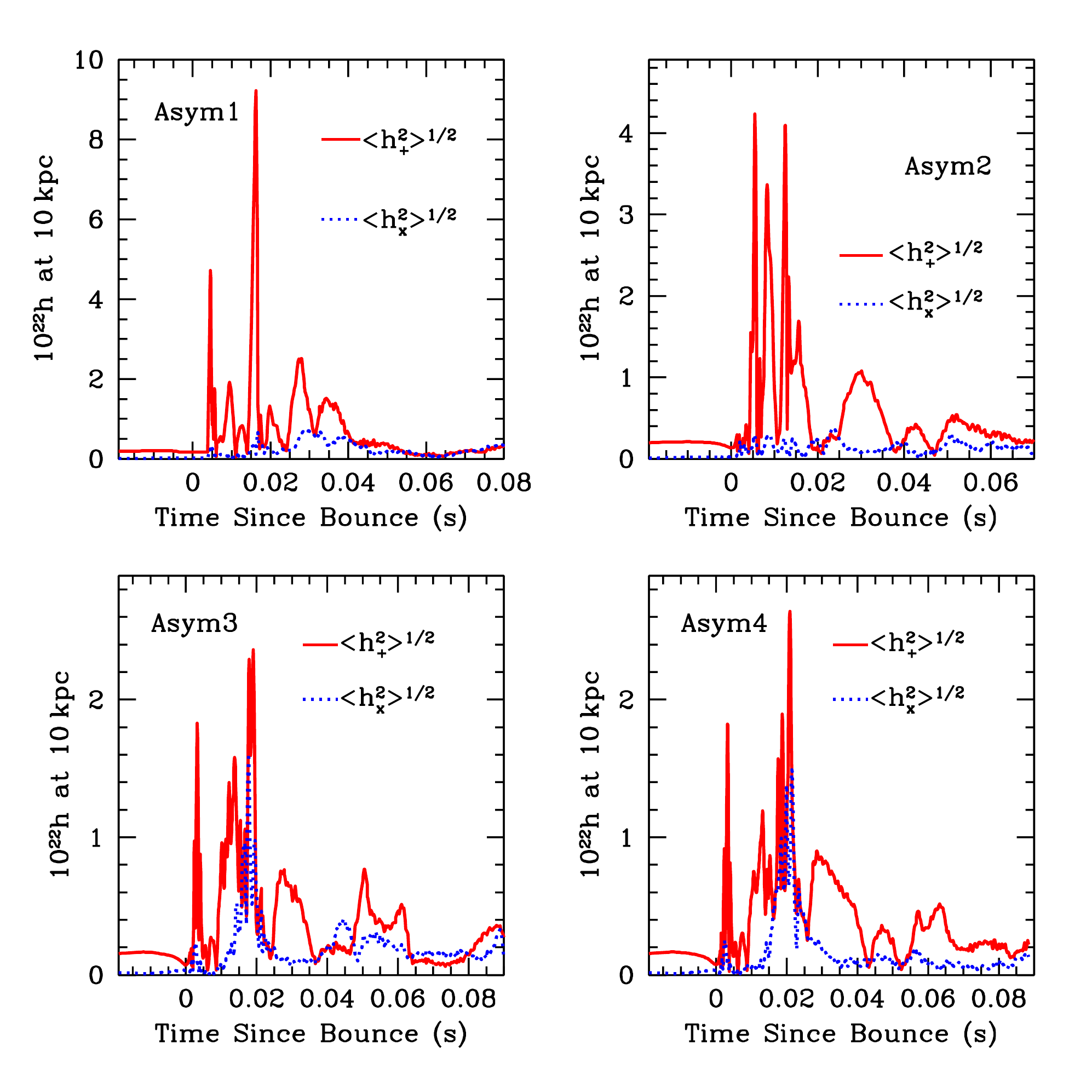}
\caption{The angle averaged wave amplitudes ($<h^2_{+}>^{1/2}$: solid
  line,$<h^2_{x}>^{1/2}$: dotted line) for the mass motions from 4
  supernovae with global perturbations prior to collapse.  Asym1
  corresponds to a 25\% global perturbation throughout the entire core
  (assuming oscillatory modes in the neutron star).  Asym2 corresponds
  to a 40\% global perturbation in the burning shells only.  Asym3 and
  Asym4 correspond to 30\% pertubations in the burning shells, with 
  and without momentum being carried away by neutrinos\cite{FHH04}.}
\label{fig:gwasym}
\end{figure}

As the rotating core collapses, rotational support develops a
quadrupole moment in the mass distribution that changes radically with
time.  Figure~\ref{fig:gwrot} shows the signal produced for a range of
rotating models, comparing fast and slow-rotating models\cite{FHH04}.
The fastest rotating cores gain enough support from centrifugal forces
to alter the peak density achieved at bounce.  By measuring the
gravitational wave signal, we can determine whether the rotation is
fast enough to alter the core densities and hence remove this
uncertainty in interpreting the neutrino signal.  Currently, the lack
of understanding of magnetic braking within stars makes it very
difficult for stellar modelers to predict the pre-collapse rotation of
stars.  Bounce signals will not only allow us to better understand the
nuclear physics, but also help constrain stellar models.  The amount
of rotation can also alter the characteristics of the explosion and
there is a synergy between diagnostics measuring the explosion
asymmetry: e.g. light curve (sec.~\ref{sec:SNLC}) and ejecta remnants
(sec.~\ref{sec:ejectaremnant}).

\begin{figure}
\includegraphics[width=8cm]{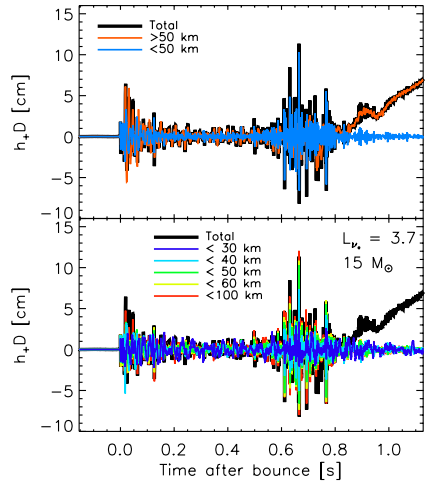}
\caption{GW waveforms, $h_+D$ vs. time, showing the contributions of
  PNS convection and low mode convection, presumably the Standing
  Shock Accretion Instability (SASI).  For reference, the entire GW
  signal is shown in both panels (solid-black line).  The signal
  originating from $> 50$ km (orange, online version) and $< 50$ km
  (blue, online version) is shown in the top panel.  This radius is
  roughly the division between nonlinear SASI motions and PNS
  convection motions.  Most, but not all, of the signal associated
  with prompt convection originates in the outer convection zone.
  There is a non-negligible contribution from the PNS.  The monotonic
  rise in the GW strain during explosion clearly originates from the
  outer (exploding) regions.  Even though the region for the
  convection/SASI and its nonlinear motions are above 50 km, these
  motions influence the PNS convective motions below 50 km.  It is
  telling that once the model explodes, and the nonlinear SASI motions
  subside but the PNS convection does not, the GW signal from below 50
  km diminishes as well.  The bottom panel shows the GW signal from
  five regions, each with different outer radii (30, 40, 50, 60, and
  100 km).  The strengthening of the GW signal associated with the
  SASI is apparent for all, suggesting that the influence of the SASI
  diminishes only gradually with depth.  Reprinted with permission
  from J.W. Murphy, C.D. Ott, A. Burrows, Astrophysical Journal,
  Vol. 707, 1173 (2009)\cite{murphy09}.}
\label{fig:gwconv}
\end{figure}

After the bounce shock stalls, convection develops both above and
below the proto-nsutron star surface.  Asymmetries in the collapse,
caused by explosive silicon burning in the progenitor just prior to
collapse, could also produce a strong gravitational wave signal.
Figure~\ref{fig:gwasym} shows the signals from asymmetric
collapse\cite{FHH04}.  Typically, large initial asymmetries are
required to produce strong signals or alter the peak bounce densities.
These signals are distinct from rotating signals and an observed
gravitational wave signal will also provide key information on the
progenitor evolution (both spin and late-stage burning phases).

Low-mode convection above the proto-neutron star produces rapidly
varying quadrupole moments and the resulting emission may also produce
a distinguishable gravitational wave signal\cite{murphy09}, providing
insight into the nature of the convection that we now believe is
important for the supernova
explosion\cite{herant94,fryer07a,blondin03}.  This low mode convection
produces a specific gravitational wave signature
(Fig.\ref{fig:gwconv}\cite{murphy09}) that can be distinguished from
higher mode convection with a sufficiently strong gravitational
signature.  As we shall see, the distribution of the compact remnant 
masses may also provide clues into this convective engine and we 
can test our models against these masses in preparation for any 
gravitational waves observations (Sec.~\ref{sec:compactremnant}).

Gravitational waves provide probes into 3 key characteristics of the
supernova explosion: stellar rotation, bounce asymmetries, and the
explosive engine.  In all cases, these have ties to the inital
progenitor and the explosive engine, and gravitational waves will
complement multiple additional diagnostics: progenitor observations
(section~\ref{sec:progenitor}), light-curves (section~\ref{sec:SNLC}),
and ejecta remnants (section~\ref{sec:ejectaremnant}).  But
gravitational waves suffer some of the same limitations as neutrino
diagnostics.  Even with the upcoming advanced detectors, gravitational
wave detections will be limited to supernova occuring in the local
group.  We have not yet detected a gravitational wave signature from
stellar collapse and, although it is extremely likely that
gravitational wave detectors will observe the mergers of binary
neutron star systems in the next 5-10 years
(Sec.~\ref{sec:compactremnant}), next generation gravitational wave
detectors may be needed to make gravitational wave detections of
core-collapse supernovae commonplace events.

\section{Progenitors}
\label{sec:progenitor}

Without clean, direct probes, the astrophysical problem becomes
increasingly complex.  We must understand our initial conditions and
use a wide variety of diagnostics to better constrain the supernova
engine.  Astronomers have used supernova rate observations
(e.g.\cite{cappellaro99, van05, li07, smartt09a}, see~\cite{smartt09b}
for a review), host galaxy environments\cite{anderson12,kelly12} and
massive star populations\cite{fryer07b} to help constrain the possible
progenitors of supernovae.  But such constraints require
interpretation from population studies and have had mixed results.
For example, one of the basic disagreements in supernova progenitors
has been whether the progenitors of type Ib/c supernovae are produced
by binaries\cite{pods92,vanbeveren07,fryer07b,vanbeveren13} or single
stars\cite{meynet11}.  Despite extensive observations of both the
relative rate of type Ib/c and type II supernovae (and the dependence
of the relative rates on the host galaxy metallicity), this remains an
open question in our understanding of supernova progenitors.
Observations of massive stars may be solving this debate,currently
arguing that over 50\% of massive stars are in binaries that will
interact in their lifetime\cite{kiminki07,kobulnicky08,kiminki12}, and
we will have to understand binary interactions to understand supernova
progenitors.

The more reliable constraint on the nature of the progenitor has been
direct observations of these progenitors.  Observing the pre-supernova
star is akin to finding a needle in a haystack.  Until large survey
archives became available, the only detected progenitors were limited
to serendipitous observations at the location of nearby supernovae.
These two events, supernova 1993J and 1987A demonstrated just how
complex the supernova could be with both being best explained by a
binary system where binary interactions shaped the progenitor and
circumstellar material, defining many of the supernova
features\cite{pods93}.  However, with the Hubble Space Telescope
archive, the era of supernova progenitor observations has begun.  As
future surveys make available their data in archives, the number 
of progenitor observations should increase dramatically.

Smartt \cite{smartt09b} provides an extensive review of the current
supernova progenitor database.  At that time, surveys had detected 8
new supernova progenitors.  From these detections (and analyzed
through stellar models), astronomers can estimate the progenitor mass.
The Smartt\cite{smartt09b} review also includes 12 upper limits that
place constraining limits on the upper star mass.
Table~\ref{table:progenitors} summarizes the results from
Smartt\cite{smartt09a,smartt09b}.  The number of progenitor
observations continues to grow and table~\ref{table:progenitors}
includes these new systems.

\begin{table}
\caption{\label{table:progenitors}}
\begin{ruledtabular}
\begin{tabular}{lc}
Name & Mass \\
\hline
Serendipitous \\
\hline
SN1987A & 14-20M$_\odot$ \\
SN1993J & $\sim15$M$_\odot$ \\
\hline
Gold Set \\
\hline
SN2003gd & 6-12M$_\odot$ \\
SN2005cs & 6-10M$_\odot$ \\
SN2008bk & 7.5-9.5M$_\odot \rightarrow$ 11.2-14.6M$_\odot$\cite{maund13}  \\
SN2004dj & 12-20M$_\odot$  \\
SN2004am & 9-19M$_\odot$  \\
\hline
Silver Set \\
\hline
SN1999ev & 15-18M$_\odot$  \\
SN2004A & 7-12M$_\odot$  \\
SN2004et & 8-14M$_\odot$  \\
\hline
Bronze Set \\
\hline
SN1999an & $<$18M$_\odot$ \\
SN1999br & $<$15M$_\odot$ \\
SN1999em & $<$15M$_\odot$ \\
SN1999ev & 12-22M$_\odot$ \\
SN1999gi & $<$14M$_\odot$ \\
SN2001du & $<$15M$_\odot$ \\
SN2002hh & $<$18M$_\odot$ \\
SN2003ie & $<$25M$_\odot$ \\
SN2004dg & $<$12M$_\odot$ \\
SN2005cs & 6-10M$_\odot$ \\
SN2006my & $<$13M$_\odot$ \\
SN2006ov & $<$10M$_\odot$ \\
SN2007aa & $<$12M$_\odot$ \\
SN2008bk & 8-12M$_\odot$ \\
\hline
New \\
\hline
SN2008ax & $10-14$M$_\odot$ or $\sim28$M$_\odot$\cite{crocket08} \\
SN2008cn & $13-17$M$_\odot$\cite{elias09} \\
SN2009md & $7-15$M$_\odot$\cite{fraser11} \\
SN2011dh & $13-22$M$_\odot$\cite{vandyk11,szczygiel12,vandyk13} \\
SN2012aw & $\sim 17-18$M$_\odot$\cite{vandyk12} \\
iPTF13bvn & $\sim$31-35M$_\odot$\cite{cao13,groh13} \\

\end{tabular}
\end{ruledtabular}
\end{table}

Typically, supernovae with observed progenitors also have detailed 
observations of the supernova outburst.  Coupled with stellar models, we can
place constraints on the inner core that then determines the evolution
of the collapse and supernova explosion.  We will discuss how these
constraints can couple with supernova light-curves to understand the
progenitor and explosion energy in Sec.\ref{sec:SNLC}.  From the table,
we see that most supernova progenitors appear to be stars below
20M$_\odot$.  This matches predictions by core-collapse
models\cite{fryer99,heger03} where strong supernovae are produced on
in stars with masses below $\sim 18-23$M$_\odot$ (massive stars can
still produce strong supernovae if they experience sufficient mass
loss to alter the core).  

Also note that roughly half of stars have masses below
$\sim$12M$_\odot$.  For initial mass functions (IMFs) lying between
2.35 and 2.7 and upper and lower limits of supernova formation
($M_{\rm SN upper}=20M_\odot, M_{\rm SN lower}=8-10M_\odot$), the
fraction of supernovae below $\sim$12M$_\odot$ is expected to lie
between 36-63\% of all stars.  As we build up the statistics on this
result, we may be able to place constraints on the lower mass limit
for supernovae and the stellar models that predict this limit.

Progenitor observations place constraints on the initial conditions
that help us take advantage of all other diagnostics.  The constraints
on the collapsing core strengthen what we can learn from neutrinos
(section~\ref{sec:neutrinos}), gravitational waves
(section~\ref{sec:GW}) as well as our theoretical models so that we
can produce more reliable explosion energies (important for all
diagnostics).  Constraints on the circumstellar medium also reduce the
uncertainties in the light-curve calculations
(section~\ref{sec:SNLC}).

But these results must be taken with some caution.  For example, new
observations of the reddening for the progenitor of SN 2008bk (one of
the gold set from Smartt et al.\cite{smartt09b}), suggests that the
progenitor was highly reddened, altering the mass prediction for the
progenitor star for this supernovae\cite{maund13}.  This demonstrates
that even our best progenitor masses in the current list are not set
in stone.  The growth of this data is placing new mandates on stellar
evolution models and, with these improvements, we will be able to
place stronger constraints on supernova progenitors.  With new surveys
and broader attempts at archiving these surveys, the rate of supernova
progenitor detection is expected to increase dramatically in the next
decade.  Coupled with supernova light-curves, these observations help
identify the initial conditions of our nuclear-physics,
astrophysics-laboratory experiment.

\section{Supernova Light-Curves}
\label{sec:SNLC}

Supernovae were first discovered by their transient optical emission
and it is not surprising that most of our data on supernovae is
concentrated in these bands.  With transient observatories (both on
ground and in space), the bandwidth for which we have coverage of
supernovae now extends from the radio up to the gamma-ray.  This
broader range, in particular the opening up of the UV and X-ray bands
from the Swift\cite{gehrels04} and now NuSTAR\cite{harrison13}
satellites, has allowed astronomers to probe the interaction of the
supernova shock with the circumstellar medium.  Recent transient 
surveys, like PanStarrs and PTF are providing a wealth of new data in 
the optical bands, including many pecuilar transients and more 
complete light curves with observations well before peak.  These 
early observations have led to a growing list of supernova observations 
covering very early times when the supernova shock breaks out of the 
stellar photosphere.  

\begin{figure}[ht]
\includegraphics[width=8cm]{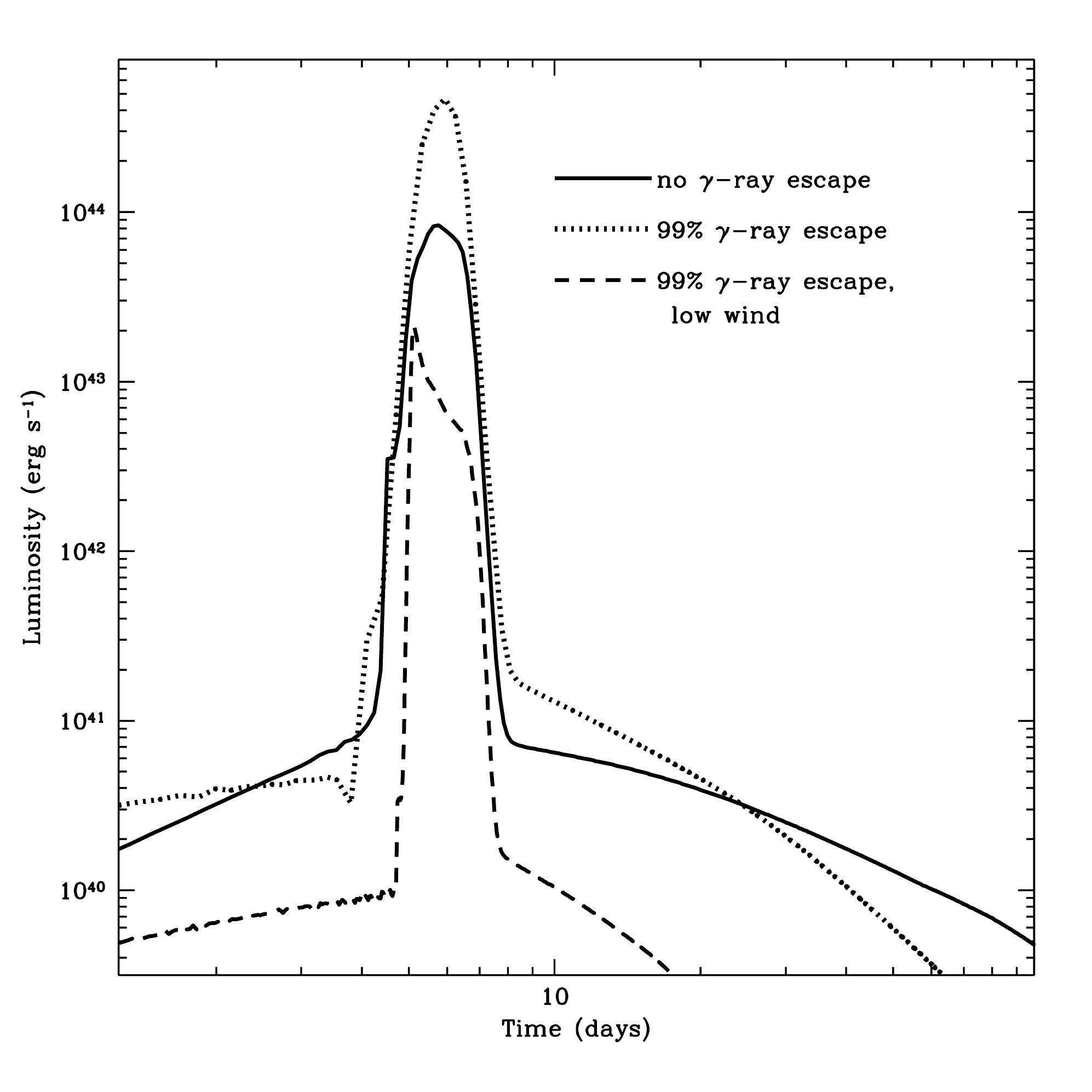}
\caption{Luminosity versus time for 3 supernova models using the same
  progenitor and explosion energy, but varying the amount of energy
  deposited from $^{56}$Ni decay\cite{fryer07c}.  If decay energy
  dominated the light-curve, by reducing the amount of energy
  deposited, we'd expect a dimmer supernova.  In fact, the opposite is
  the case.  In this study\cite{fryer07c}, the focus is on the energy
  lost because the $\gamma$-rays did not interact with the stellar
  ejecta, but this is equivalent to lowering the total $^{56}$Ni in
  the explosion.  However we do see that, because shock heating
  dominates, the amount of wind material does change the
  light-curve. }
\label{fig:lcni}
\end{figure}

Many studies of supernovae have focused on using the peak brightness
to infer the explosion energy, progenitor mass and $^{56}$Ni yield.
For many type Ia supernova progenitors, where we have already
constrained the explosion to a near-Chandrasekhar massed white dwarf
with low-density surroundings, such interpretations may be successful.
But for the massive stars producing core-collapse supernovae, mass
loss (via winds, explosive ejections or binary interactions), the
circumstellar medium can drastically alter the light curve.  In
massive stars, the peak of the light-curve may have nothing to do with
the yield of $^{56}$Ni yield.  Figure~\ref{fig:lcni} shows 3 different
light curves for a GRB-associated supernova, where the amount of
energy deposited by $^{56}$Ni decay is varied by a factor of
100\cite{fryer07c}.  This is equivalent to reducing the amount of
$^{56}$Ni produced without changing any other aspect of the explosion.
In this case, the peak light-curve is brighter with less energy
deposition from $^{56}$Ni decay.  This is because shock heating
dominates the light curve at peak.

\begin{figure}
\includegraphics[width=8cm]{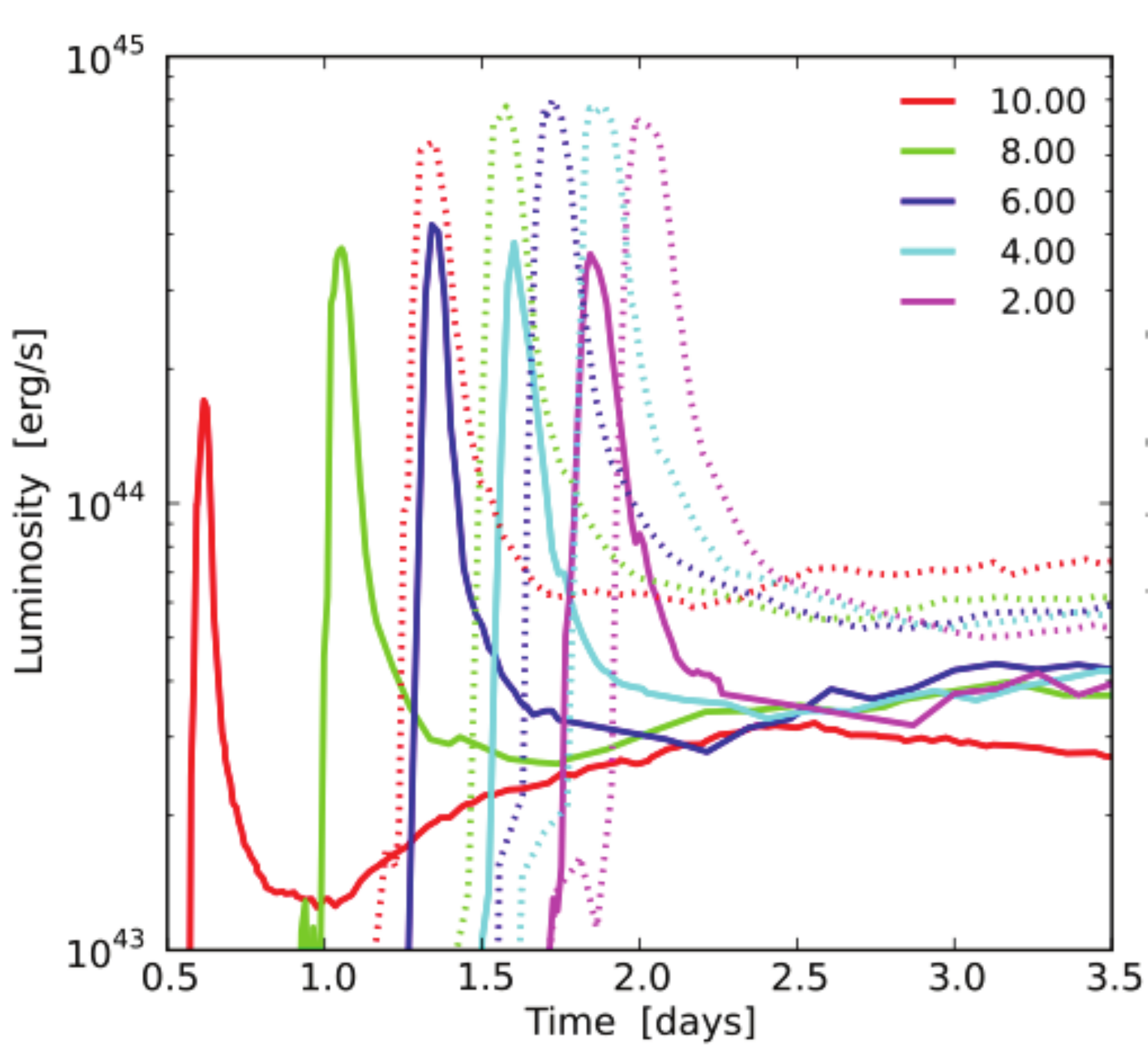}
\caption{Bolometric light-curves for shock breakout for a suite of 
models.  The first set (solid lines) shows the breakout for 5 models, 
each removing a different amount of mass from the hydrogen envelope.  
For these models, the outer layers of the star are removed in the 
mass and the mass removal reduces both the envelope mass and the 
star's radius.  In the latter set (dotted lines), the density of 
the envelope is lowered, removing mass but not removing radius.  
Although there is some degeneracy in these model results, in general 
the mass removal that also reduces the radius has a more drastic 
effect on the breakout time and duration of breakout\cite{bayless14}.}
\label{fig:lcsb}
\end{figure}

Realizing that shock heating is an important contributor to the
light-curves of many supernovae alters what we can potentially learn
from these supernovae.  Instead of measuring the amount of $^{56}$Ni
produced, observations of supernovae are more sensitive to, and can
help constrain, the circumstellar medium.  For massive stars, the
radius where the star ends is difficult to measure as the photosphere
is often in the wind or other circumstellar ejecta.  The supernova
light-curve provides extensive information about this circumstellar
medium and the winds/outbursts from stars prior to their collapse.
One of the most clear-cut ways to measure this photosphere is through
shock breakout.  The duration and peak of the shock breakout can
measure the radius of the photosphere.  For example,
figure~\ref{fig:lcsb} shows the shock breakout light-curves for a
single progenitor and explosion where the mass of the hydrogen
envelope has been reduced in two ways: removing the top layers of the
hydrogen envelop (in this manner, the radius decreases with the
decrease in mass) and reducing the envelope density (in this case, the
photospheric radius decreases only slightly with decreasing mass).
The time of the shock breakout is most sensitive to the stellar
radius, but the density in the giant can also alter the breakout time.
Without a neutrino signal, it may be difficult to pinpoint the exact
explosion time, so the width of the shock breakout will play a
stronger role in constraining the progenitor characteristics.  As
surveys begin to probe shorter and shorter transients, shock breakout
observations will become increasingly common and these probes of the
stellar photosphere will become strong constraints on the supernova
progenitor.

\begin{figure}
\includegraphics[width=8cm]{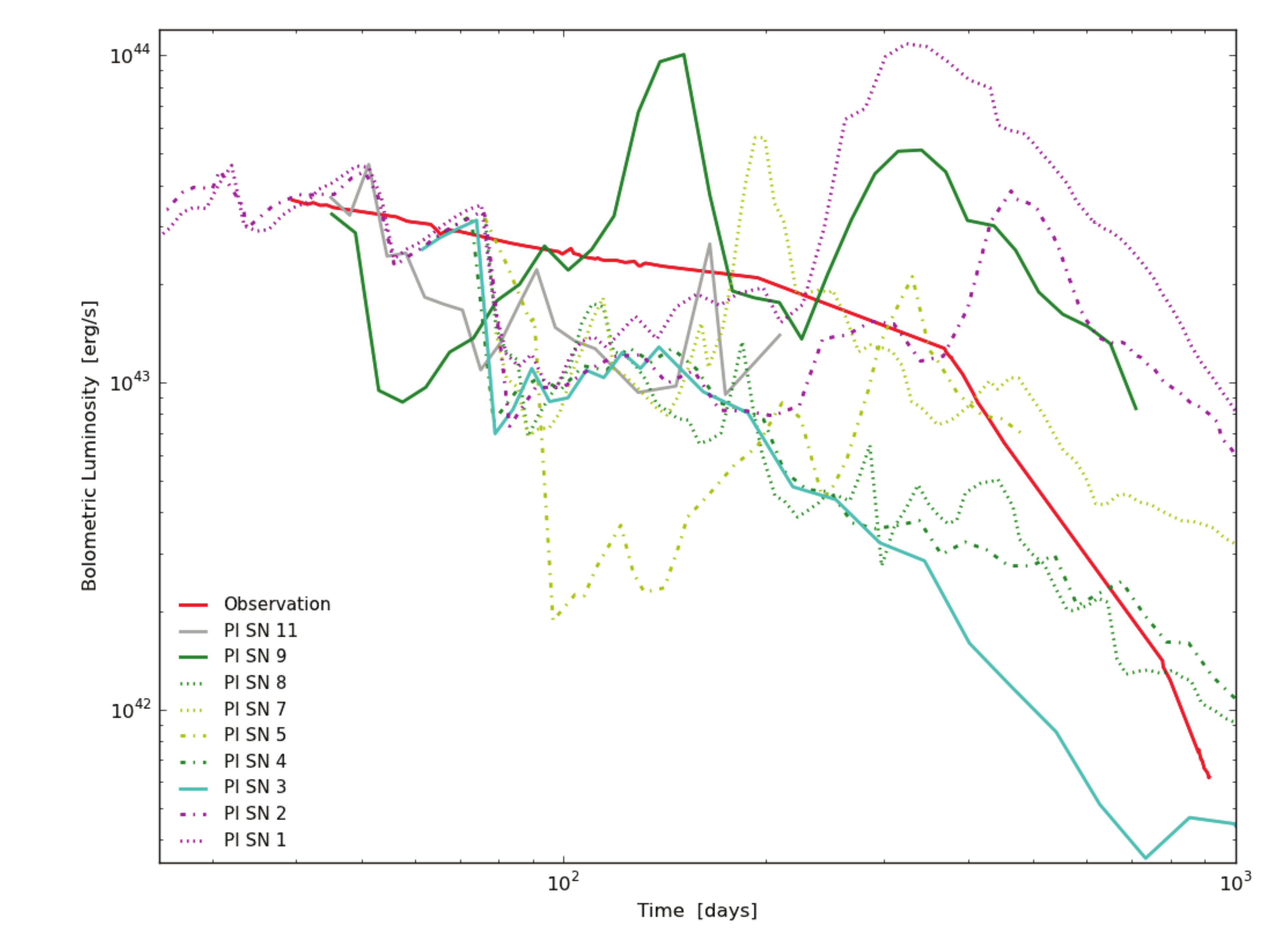}
\caption{The light curve in r-band of the SN2010jl
  supernova\cite{ofek13} coupled with a series of models striving the
  replicate this effect.  These models include a range of
  pair-instability models, although hypernova models could also be
  designed to match this data.  In any event, all models required
  considerable envelope mass and explosion energy to explain the
  extended light curve.  We expect high mass loss in both
  pair-instability and hypernova models so both the energy
  requirements and circumstellar mass point to these massive
  explosions.}
\label{fig:lc2010}
\end{figure}

Post breakout Supernova light-curves and spectra further probe the
circumstellar medium, studying the outbursts and wind ejecta from a
supernova progenitor prior to collapse.  An extreme example of this is
SN 2010jl.  In this supernova, interaction with nearly 10\,M$_\odot$
of circumstellar medium kept the supernova bright (even in the X-rays)
for over a year after the explosion\cite{ofek13}.  Observations of the
light-curve allowed astronomers to probe the characteristics of this
ejecta, arguing for extreme mass loss immediately prior to the
explosion and extremely energetic explosions (either a hypernova or
pair-instability supernova).  Figure~\ref{fig:lc2010} shows a suite of
models with the same progenitor and explosion characteristics and
varying only the details of the massive circumstellar medium.  With
these models, we find that SN2010jl can only be explained using a
large amount of circumstellar material (over a few M$_\odot$) and an
energetic explosion (either a pair-instable supernova or hypernova).
If this is a pair-instability supernova, the surrounding ejecta
demonstrates that pair-pulsations (explosive outburst produced when
pair production causes an implosion and resulting explosive burning
phase in a star) can occur before the final outburst.  Although there
is some evidence supporting the existence of these theoretically-predicted
pair-instability puslations and explosions in other observations, this
might be the strongest evidence for pair-pulsations and explosions in
our current data suite.  In this case, the supernova light-curve could
place constraints on both the supernova energy and surrounding medium.

Late-time observations of the $^{56}$Ni decay curve will still provide
insight into the total $^{56}$Ni yield.  With an adequately developed
thoery, the wide breadth of supernova light-curve and spectra data
that spans from radio up through gamma-rays can be used to not only
probe the circumstellar medium and supernova progenitor, but also the
characteristics of the supernova explosion (energy and, ultimately
explosion asymmetry).  

Theory advances are needed to take advantage of this data.  Modeling
core-collapse light-curves ultimately will require full
radiation-hydrodynamics calculations.  Solving this correctly requires
a 6 dimensional calculation modeling the energy structure, angular
distribution and spatial distribution of the photons.  To date, most
radiation hydrodynamics calculations use simplified transport
algorithms (flux-limited diffusion with gray or few
groups\cite{blinnikov93,frey13}) in 1-dimension although higher-order
transport to follow the angular distribution is technically feasible,
it is typically too costly to be implemented in full parameter
studies\cite{frey13}.  Those calculations with higher-order transport
typically do not couple the radiation to the
hydrodynamics\cite{maeda06,kasen08}.  On top of this, the treatment of
opacities often assumes local thermodynamic equilibrium, when the
atomic physics suggest that out-of-local thermodynamic equilibrium,
perhaps even non-steady state physics is important.  Again, codes
modeling this physics more accurately tend to not model
radiation-hydrodynamics\cite{mazzali93}.  Research in atomic physics
is also important.

These physics uncertainties and numerical obstacles might make it seem
that using supernova light-curves and spectra to constrain the
supernova engine is a daunting affair.  But the rise in data and
computational resources will probably make this diagnostic one of the
most powerful in the next decade.  Coupled with observations of
the progenitor as an independent diagnostic, supernovae are ideally
suited to test the transport and atomic physics theory.  Here, the
astrophysics ``experimental'' community and the high-energy density
laboratory community can be very complimentary to each other.
Laboratory experiments will also provide insight, ultimately pinning
down this physics\cite{remington06}.  As a transport physics experiment
alone, supernovae are quite useful.  But as this physics is solved,
supernova light-curves will also be a strong constraint in
understanding supernova progenitors and the supernova explosion.  With 
these constraints, we can approach our studies of nuclear physics with 
a better understanding of our supernova experiment.

\section{Compact Remnants}
\label{sec:compactremnant}

Compact remnant masses place constraints both on the equation of state
directly and on the nautre of the supernovae explosions mechanism.
Equations of state make predictions for the maximum mass of a neutron
star.  The observed maximum mass places a lower limit on this maximum
neutron star mass.  The measurement of a 2\,M$_\odot$ neutron
star\cite{demorest10} has already placed strong constraints on the
nuclear equation of state\cite{hebeler13}.  If this maximum mass
pushes upwards either through a new observation or firmer constraints
on existing neutron star mass estimates, severe constraints can be
placed on the nuclear equation of state at both sub- and supranuclear
densities.

The mass of the compact remnant can also provide clues into the
explosion mechanism.  By using orbital characteristics, the masses of
compact remnants in binaries are fairly well constrained and the mass
distributions of compact remnants have been estimated by a number of
groups\cite{finn94,bailyn98,thorsett99,ozel10,farr11,ozel12}.  Initial
mass distributions argued for bimodal, nearly delta-function, peaks.
With time, some of these peaks have filled in, but a gap in the masses
between the most massive neutron star ($\sim 2$M$_\odot$) and the the
lowest mass black hole ($\sim 5$M$_\odot$)
persists\cite{bailyn98,ozel10,farr11}.  However, note that it has been
argued that the mass gap could be an artifact in deriving orbital
inclination angles of black hole X-ray binaries, which were used in
determining the mass functions of black holes\cite{kreidberg12}.  Here
we review how the exact nature of this gap provides clues into the
supernova engine.

The observed compact remnant masses are determined by 3 factors: the
mass at explosion (depending on the explosion timescale), the amount
of fallback after the launch of the shock (dependent on the explosion
energy and asymmetry), and long-term accretion from binary companions.
For many cases, the binary accretion is small or can be
estimated\cite{willems05,fragos09}.  In such cases, the compact
remnant mass distribution can be used to constrain the nature of the
supernova explosion.

To discuss this in more detail, let's review the aspects of the
supernova explosion mechanism that affect the compact remnant mass.
The bounce shock moves out through the star until neutrino cooling,
and to a lesser extent, dissociation of the infalling silicon layer,
sap its energy and it stalls.  The region between the proto-neutron
star surface and the stalled shock is susceptible to a number of
instabilities.  Much of the early work focused on Rayleigh-Taylor
instabilities in this stalled shock
region\cite{herant94,burrows95,janka96,fryer02,fryer04}.  If these are
present, they tend to dominate.  If the Rayleigh-Taylor instability is
weak, additional instabilities can develop and dominate.  Early papers
invoking the ``Standing Accretion Shock Instability'' focused on new
instabilities, e.g. the advective-acoustic
instability\cite{blondin03,blondin06,bruenn09,hanke13}.

The Compact remnant mass distribution is able to provide insight into
which instability dominates, and therefore determine the exact
conditions of the explosion.  The advective-acoustic instability has a
growth time that is $>10$ times longer than Rayleigh-Taylor.  This
longer growth time tends to produce delayed explosions taking $\sim
0.5-1$\,s to develop\cite{blondin03}, in contrast to the 100-200\,ms
explosions produced in simulations with strong Rayleigh-Taylor
convection\cite{fryer99}.  Measuring the delay could potentially constrain 
the explosion engine\cite{fryer12}.

\begin{figure}
\includegraphics[width=8cm]{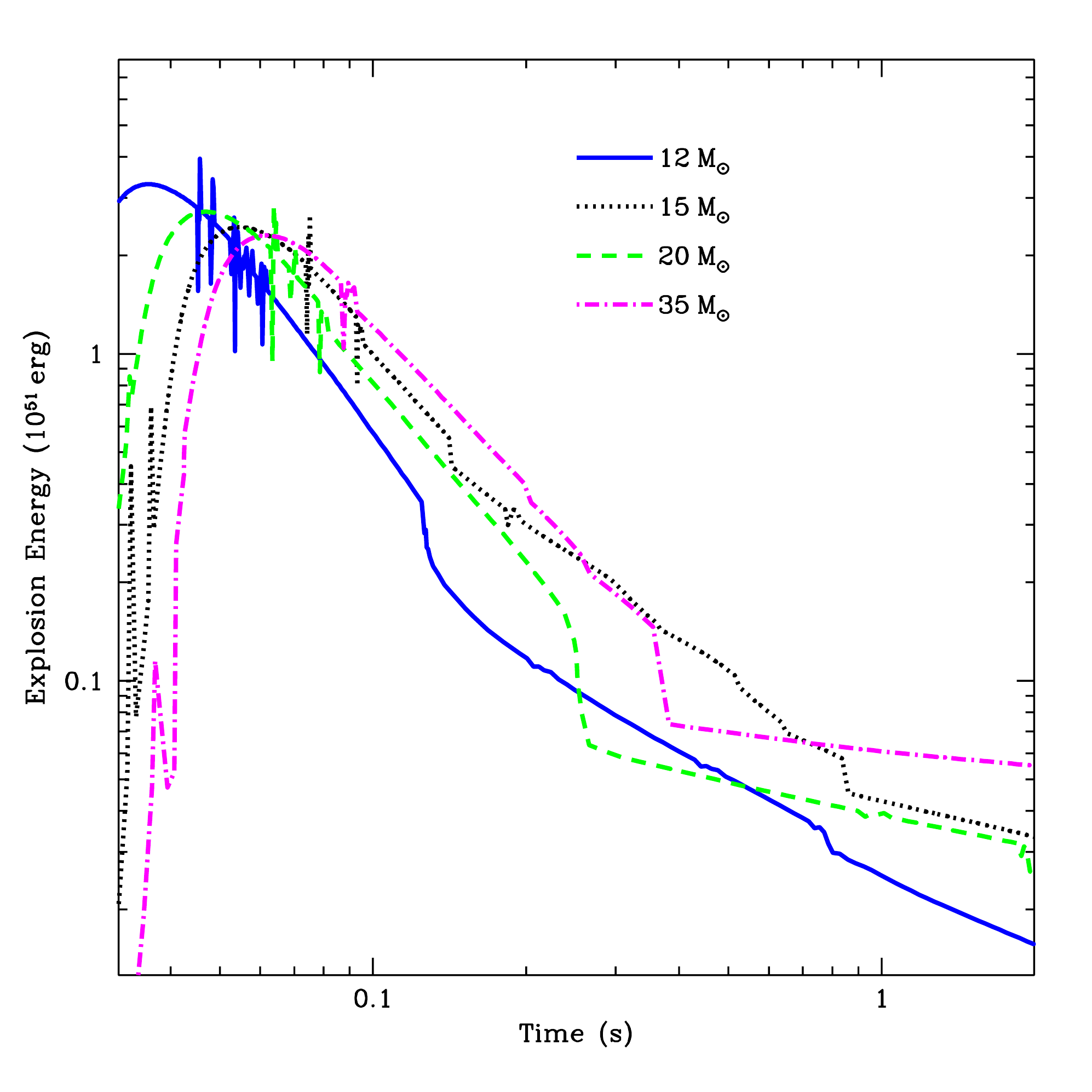}
\caption{Explosion energy as a function of time for 3 different
  progenitors.  Here we have assumed that the explosion energy is
  dominated by the the energy stored in the convective region at the
  time of the launch of the explosion.  The energy stored depends upon
  the rate of infall onto the the convection region.  This infall acts
  as a pressure-cooker lid, stronger lids (with higher infall rates)
  can store more energy in the convection region
  (see\cite{fryer06a,fryer12} for details.}
\label{fig:snexp}
\end{figure}

The duration of the explosion is not just important in determining the
neutrino signal.  A sufficiently detailed neutrino signal could both
measure the explosion duration and neutrino physics effects.  But it
is more likely that the next neutrino measurement will be pushing the
limits of detectability and the signal will be limited to just a
handful of detections.  For this, having an alternate measurement of
the delay will help us take full advantage of the neutrino signal.  If
we assume that the convection region dominates the explosion energy, a
good estimate of the explosion energy can be made by setting the
energy stored in this region when the explosion occurs and the shock
is launched.  This energy, in turn, depends upon the accretion rate of
the infalling star onto this convective region\cite{fryer06a,fryer12}.
With these assumptions, we can estimate the explosion energy as a
function of time, shown in figure~\ref{fig:snexp}.

\begin{figure}
\includegraphics[width=8cm]{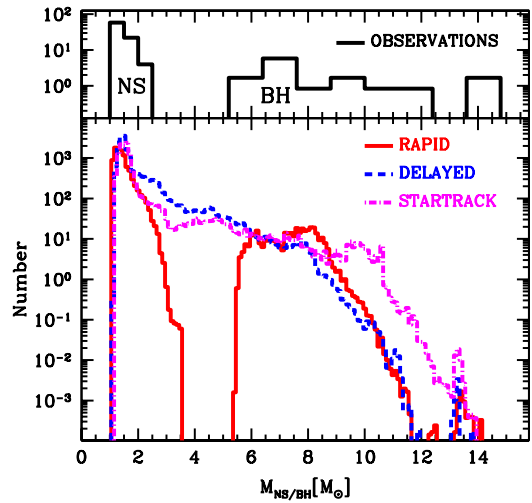}
\caption{Predicted neutron star/black hole mass distribution for the
  Galactic population of Roche lobe overflow and wind fed X-ray
  binaries.  This result employed a population synthesis method to
  generate the binaries with two supernova explosion models. In the
  model in which the explosion is driven on a rapid timescale, we note
  the same striking gap in remnant mass (almost no compact objects in
  the gap region) as found in the observations.  In contrast, the
  model in which the supernova explosion is significantly delayed
  shows a continuous compact object mass distribution. For comparison,
  we also show the remnant mass distribution commonly used in modern
  studies of Galactic and extra-galactic binaries with compact objects
  (see\cite{belczynski12} for details.)}
\label{fig:snrem}
\end{figure}

Assuming a fraction of the explosion energy goes toward unbinding the
star, one can estimate the amount of fallback in the system,
predicting a final remnant mass.  In this manner, we can estimate the
compact remnant mass as a function of initial stellar
mass\cite{fryer12}.  Including population synthesis calculations, one
can estimate the mass distribution of neutron stars and black holes in
binaries and compare them to the observed remnant mass distribution
(Fig~\ref{fig:snrem}\cite{belczynski12}).  Current studies argue that if
a gap exists in the remnant mass distribution, delayed explosions are
rare, meaning that most collapsing stars either explode quickly and
strongly or fail (or are extremely weak).  If true, this result
assures us that the neutrino signal from stellar collapse will cleanly
probe nuclear and neutrino physics and not be significantly hampered
by late-time explosion evolution or fallback.

This picture depends both on our understanding of the observations
(which may stay have biases) and theory.  We have already discussed
the observational uncertainties.  An alternate way to measure compact
remnant masses is through gravitational wave observations of the
merger of two compact remnants.  As two compact remnants merge, they
produce a strong gravitational wave signal.  Compact mergers are
expected to be the primary gravitational wave source for the advanced
LIGO detector with a predicted rate of a few to one hundred detections
per year when the detector is at full
operations\cite{abadie10,dominik12}.  To leading order, the phasing of
the gravitational wave signal from these mergers depends upon the
chirp mass of the binary: $\mu = (m_1m_2)^{3/5}(m_1+m_2)^{-1/5}$,
where $m_1$ and $m_2$ are the masses of the two compact remnant masses
in the binary.  Higher order terms in the signal rely on mass ratios
and, in principle, the masses of both compact remnants could be
determined from a strong gravitational wave detection.  However, in
practice, it is difficult to obtain exact masses with the typical
signal expected with advanced LIGO\cite{hannam13}.  Nonetheless, these
detections should be able to provide independent information on the
mass gap.

For our theoretical interpretation,
we have made a series of assumptions.  First, we have assumed that the
explosion energy is stored in the convective region.  If another
energy source, for example rotational energy in the core, can be
tapped, estimates using only the energy in the convective region
underestimate the total explosion energy.  These additional energy
sources might make it that a delayed explosion can still produce a
mass gap.  We will have to use other means (supernova light curves) to
determine whether such additional energy sources are present in the
explosion.  Other theoretical errors include our understanding of
fallback.  We will discuss this further in section~\ref{sec:postexp}.

\section{Supernova Remnant Ejecta}
\label{sec:ejectaremnant}

The term supernova remnant is also used to refer to the interaction of
the ejecta from the supernova explosion with the surrounding
circumstellar medium.  A series of analytic derivations are used to
estimate the explosion energy, stellar structure, and the nature of
the interstellar medium using supernova remnants\cite{chevalier82a,chevalier82b}.
For most remnants, these estimates provide a first estimate of the
initial supernova conditions.  However, these analytic derivations
tend to simplify the physics too much to be used to strongly constrain
the supernova engine.  In addition, supernova remnants are observed
well after the actual supernova event and, for the most part, ejecta 
remnant observations can not be easily coupled with the other diagnostics 
of supernovae (the light-echo observation of the supernova outburst for the 
Cassiopeia A remnant being the one counter-example\cite{krause05}).  But supernova 
ejecta remnants provide insight into the explosion mechanism not detected 
by other diagnostic probes and despite the limitations of these observations, 
ejecta remnants remain powerful probes in our understanding of supernovae.

Until recently, most of the studies of supernova remnants have focused
on hydrodynamics\cite{chevalier92,blondin96} and particle
acceleration\cite{blandford78,bell78a,bell78b,jones91,ellison00}, for
a review, see\cite{reynolds08}.  With advent of spatially resolved
cosmic ray detectors observations of supernova
remnants(e.g.\cite{aharonian05,aliu13}), studies of particle
acceleration have intensified
(e.g. \cite{yuan11,telezhinsky13,morlino13,bykov13}.  Coupled with the
wealth of observations from radio (e.g. VLBA) through gamma-rays
(e.g. Fermi, HESS\cite{aharonian13}), we are building a strong picture
of particle acceleration processes.  Studies of hydrodynamics have
also taken advantage of the growing, time-resolved data from the
Chandra satellite (e.g.\cite{hwang12}) of the development of
instabilities and the ability to calculate these instabilities in
3-dimensions\cite{obergaulinger13,ellinger13}.  With detailed models
and a growing understanding of the initial conditions (see below),
this wealth of data provides an ideal setting to study turbulence
models.

Recent work, both observational and theoretical, has shown the
potential of the supernova remnant as a powerful diagnostic tool to
study the supernova explosion mechanism. Here we focus on one of the
best-studied remnants: the Cassiopeia A supernova remnant.

First we review the theoretical advances in modeling the Cassiopeia A
supernova remnant.  Progenitor estimates for Cassiopeia A have ranged
from a 16\,M$_\odot$ single star\cite{chevalier03} to a very massive
progenitor 30-60\,M$_\odot$\cite{fesen91}.  These efforts based their
conclusions using stellar models in the literature.  Recently,
theorists have modeled Cassiopeia A progneitors, focusing on the
constraints placed by it observations.  This includes not only the
shock position and velocity, but the nucleosynthetic yields and the
presence of nitrogen-rich knots (allowing stellar modellers to predict
that most of the hydrogen was removed prior to collapse, but most of
the helium core must be retained)\cite{young06}.

The observational study of Cassiopeia A extends across many decades in
energy. The rapidly moving infra-red features in the region
surrounding Cassiopeia A have been interpreted by observers as the
light echo from the supernova explosion (whose light originally
arrived to Earth 330 years ago) that formed the Cassiopeia A
remnant\cite{krause05}. With this interpretation, we now have an
integrated spectrum of this supernovae that can also place constraints
on its progenitor as that of a core-collapse Type IIb supernova with
characteristics similar to that of SN 1993J.

\begin{figure}
\includegraphics[width=8cm]{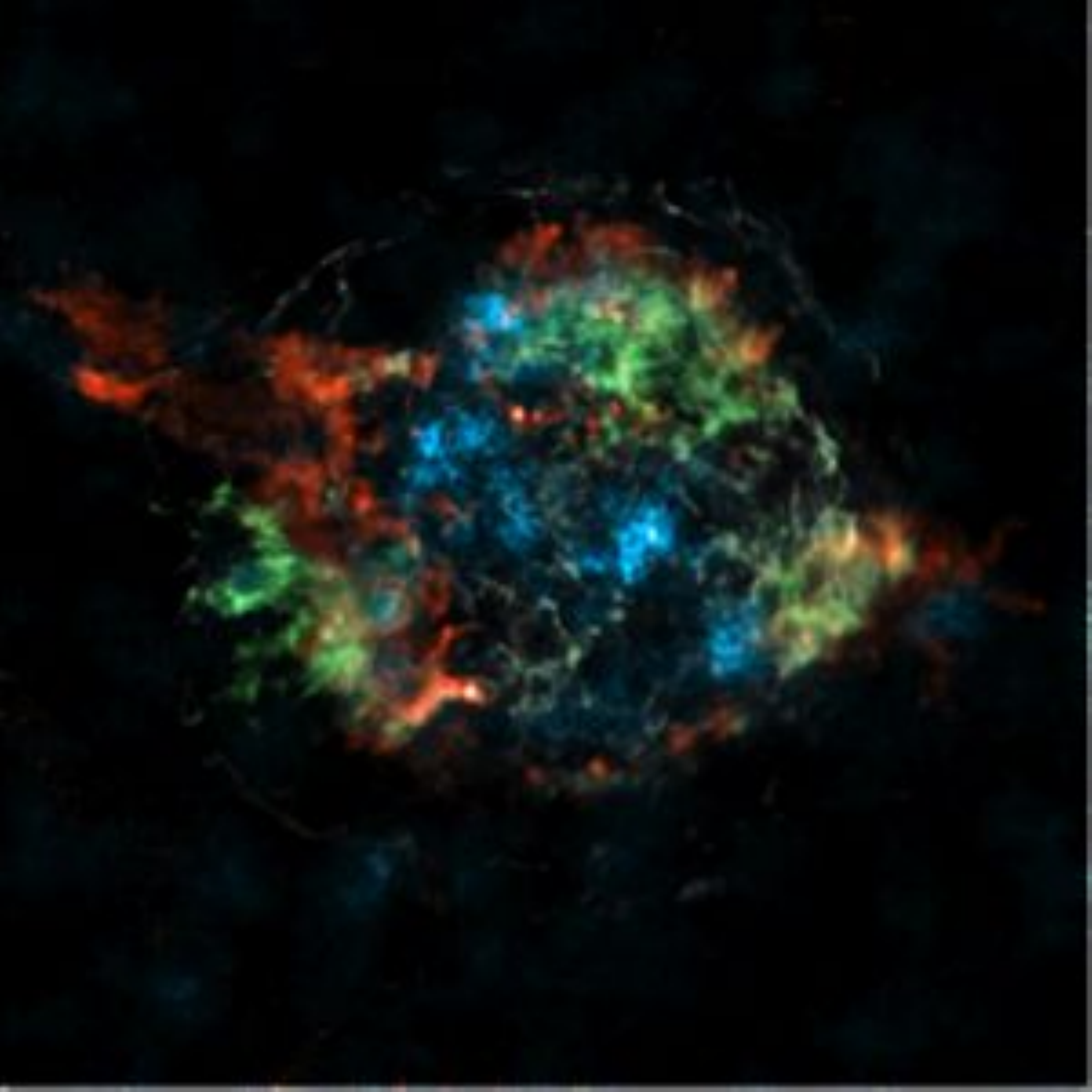}
\caption{A image showing a composite of the continuum (red), iron line
  emission (green) and $^{44}$Ti-decay line distribution for the
  Cassiopeia A supernova
  remnant\cite{grefenstette13}.\label{fig:nustar}}
\end{figure}

Cassiopeia A also demonstrates the danger in over-interpreting the
spatial distriubtion of the remnant ejecta seen in soft X-rays by
chandra and XMM.  Until the launch of the NuSTAR (Nuclear
Spectroscopic Telescope Array) satellite\cite{harrison13}, our view of
remnants was limited to the material that is currently shocked to
sufficient temperatures to be observed.  For Cassiopeia A, maps of
this shocked gas (Figure~\ref{fig:nustar}) argued both for a highly
asymmetric (almost jet-like) explosion where somehow the iron was
mixed out extensively (beyond the silicon layer).  However, the NuSTAR
satellite opened up a new window into studying these remnants.  The
3-80keV bandwidth of NuSTAR includes the 67.86 and 78.36 keV lines
produced in the decay of $^{44}$Ti, producing a map of all the
$^{44}$Ti produced in the supernova shock, not just the shocked
$^{44}$Ti.  Because $^{44}$Ti is primarily produced with the same
inner ejecta where $^{56}$Ni (which decays to iron) is produced, these
direct observations of $^{44}$Ti provide an ideal tracers of the inner
engine.  The NuSTAR observations (Figure~\ref{fig:nustar}) demonstrate
just how misleading shock-limited observations of remnant ejecta can
be: unless some unkown mechanism decouples the $^{44}$Ti and $^{56}$Ni
spatial distributions, most of the iron in Cassiopeia A remains
unshocked and hidden.  The puzzle of the bizzare iron distribution in
Cassiopeia A was simply caused by only the shocked material.  With
NuSTAR, we not only construct a picture of a ``normal'' supernova, but
we might have the first real evidence that the low-mode supernova
engine model (either caused by low-mode Rayleigh-Taylor convection or
advective-acoustic instabilities) is behind the Cassiopeia A
explosion.

Observations of $^{44}$Ti open up a new realm in studying
nucleosynthetic yields and nuclear physics in supernovae.  $^{44}$Ti
is produced primarily by material shocked to near NSE temperatures ($>
4-5\times10^9$K).  But the exact amount of $^{44}$Ti produced depends
sensitively on the conditions of the material as it expands and cools.  
Indeed, even the way the $^{44}$Ti is produced changes depending on 
the exact details\cite{magkotsios10}.  An in depth review of these paths
(identifying 4 distinct paths), and their dependence on the
uncertainties in nuclear cross-sections exists\cite{magkotsios10} and
we will simply summarize some of the critical points here.  The
separate paths include high density regions where the interaction rate
for $\alpha$ particles is so high that very little helium is left in
the final distribution after expansion.  In these conditions,
quasi-statistical equilibrium values provide an accurate description
of the final yields and these yields are primarily dependent on masses
and Q-values.  At lower densities, the isotopes cluster and the Si-Ca
cluster (where $^{44}$Ti lies) flows into the iron-peak, producing a
chasm in $^{44}$Ti production.  Other regions probe peculiarities in
the $\alpha$-rich freeze-out.  

\begin{figure}
\includegraphics[width=8cm]{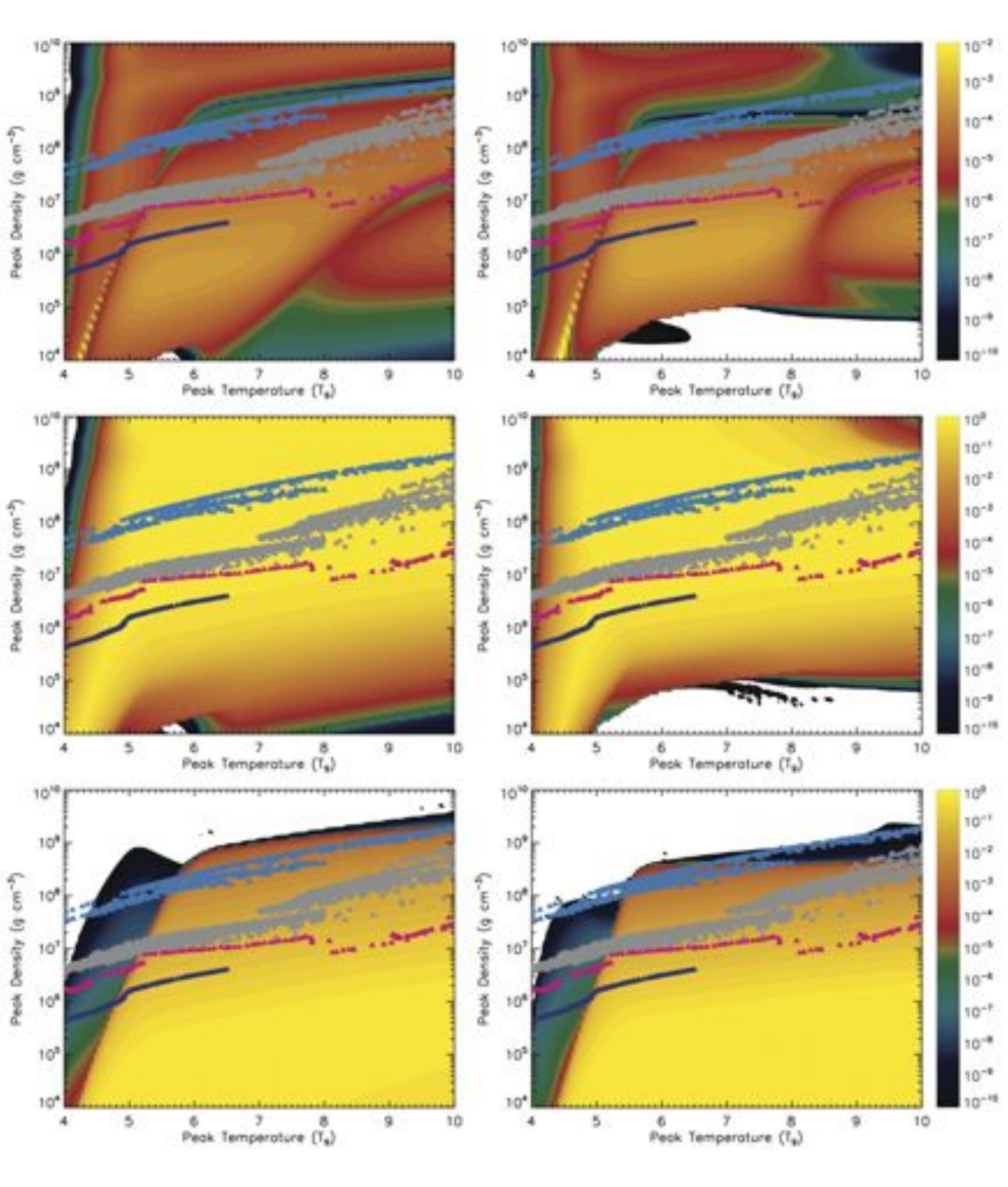}
\caption{Final mass fraction of $^{44}$Ti (first row), $^{56}$Ni
  (second row), and $^{4}$He (third row) in the peak
  temperature-density plane.  The left column assumes an exponential
  decay evolution for the density and temperature evolution with time
  whereas the right column assumes a power-law pevolutio
  profile\cite{magkotsios10}.  The different colored triangles refer
  to the trajectories of different supernova and hypernova explosion
  models - blue for a one-dimensional Cassiopeia A
  model\cite{young06}, gray for the two-dimensional explosion of
  rotating progrenitor E15B\cite{fryer00}, pink for a one-dimensional
  hypernova model\cite{fryer06b} and cyan for a two-dimensional
  magnetohydrodynmaic collapsar model.  The exact location of the
  valley in $^{44}$Ti production is extremely sensitive to the details
  in the nuclear physics.
\label{fig:magkotsios}}
\end{figure}

Observing the distribution of $^{44}$Ti yields and coupled with our
understanding of nuclear reaction rates, we can probe the exact
details of the supernova engine.  For example, the chasm in $^{44}$Ti
yield is sensitive to details in the nuclear cross-sections.
Figure~\ref{fig:magkotsios}  shows the titanium
yield for a grid of temperatures and densities at peak temperature.
The chasm is quite narrow as it moves up in density and temperature
until it the temperature reaches $6\times10^9$K, where it then
broadens and extends to a temperature of $10^{10}$K at a density of a
few times $10^8\,{\rm g cm^{-3}}$.  The actual density where the chasm
flattens out depends upon details in the nuclear physics.  On top of
this plot are the particles from a series of supernova explosions.  In
particular, note that the the data from the 1-dimensional Cassiopeia A
explosions model lies right near the chasm.  In such scenarios, the
final $^{44}$Ti yield is extremely sensitive to the nuclear physics
determining th position of the chasm.  If we had other constraints on
the explosion trajectories, the $^{44}$Ti yield becomes a strong
nuclear physics probe.

The ejecta remnants of supernovae are ideally suited to probing
nucleosynthetic yields.  These yields can be used to both study
nuclear reaction rates as well as better constrain the supernova
engine.  In particular, ejecta remnants have the potential to provide
strong constraints on the asymmetry in the supernova explosion.
Unfortunately, in most cases, we will not be able to combine disparate
observations to place strong constraints on the supernova explosion,
with the exception of a few examples such as Casseiopeia A (where we
can observe the light-echo supernova) or SN 1987A (which is young
enough and near enough to us so that the next generation of imaging
hard X-ray satellites will resolve the quickly growing SN 1987A
remnant).  But ejecta remnant observations alone will still provide
insight into the asymmetries and nucleosynthesis in supernovae that
can be incorporated into our current understanding of supernova
explosions.

\section{Nucleosynthetic yields:  Stars and Meteorites}
\label{sec:nucobs}

Nucleosynthetic yields of key elements like $^{56}$Ni and $^{44}$Ti
can be measured in individual supernovae in the outburst itself or in
the late-time supernova remnant, but more detailed yields are typically 
observed only in an integrated sense, either in stars after the supernova 
ejecta has been swept up in a new burst of star formation or in meteorites.  
Although these observations are limited to integrated yields that might include 
components from multiple supernovae, they have been extensively used to constrain 
our understanding of the supernova explosion.  Here we review just a few 
examples of this diagnostic is used in practice.

Extremely metal poor stars are ideal probes of the nucleosynthetic
yields of the first high-mass, zero metallicity stars.  Produced in
the early universe, these low-mass stars that survive to the present
day are being observed with increasing frequency.  The first
surveys showed the success of such work in the early 80s and
90s\cite{beers85,ryan91,beers92}, but new surveys have rapidly
accelerated this work (e.g.\cite{aoki13,garcia13,lee13,placco13}).  A
wide variety of surveys (to name a few, SDSS/SEGUE\cite{aoki13},
APOGEE\cite{garcia13}, LAMOST\cite{deng12}) are being conducted or
planned that will increase this data significantly in the next decade.
Astronomers can also use abundance patterns in damped Ly$\alpha$
systems to study the nucleosnthetic yields of the early universe\cite{fenner04}.
These studies, focusing on the second generation of stars, have the
advantage that they are not contaminated by too many supernova explosions.  

Studying the yields of the earliest stars can provide insight into the
first generation of stars and their formation.  This first generation
of star formation might produce a more exotic set of stars that can
provide deeper insight into stellar evolution and supernovae
(e.g. pair-instability supernovae).  The explosions of these systems
make firm predictions on the relative abundances of odd and even-Z
elements\cite{nakamura99,heger02} and observations of this odd-even
effect\cite{ryan96,fenner04} could provide deep insight into stellar
evolution.

The origin of the r-process is another feature of supernovae that has
been studied using abundance patterns in low-metallicity stars.  With
these observations, scientists have argued that at least some of the
r-process must be produced in massive stars\cite{qian02} and
scientists have gone so far to as to use these observations to justify
particular r-process sites\cite{argast04,qian12}.  Although there is
still quite a bit of uncertainty in these observations, these
constraints will ultimately tell us about both the supernova
progenitor and its subsequent explosion.  We will discuss this in more
detail in section~\ref{sec:postexp}.

For high-z elements, it becomes difficult to determine the relative 
isotopic abundances in the spectra as the line differences become difficult 
to distinguish.  However, in meteoritic data, the relative isotopic abundances 
can be determined (for a review, see\cite{anders89}).  These isotopic abundances 
can be used to constrain new nuclear/neutrino physics abundances.  For example, 
neutrinos can alter the nucleosynthetic yields of supernovae.   Neutrinos 
can excite elements, even unbinding some of the particles.  The evaporation of 
a neutron or proton can lead to new nucleosynthetic paths, altering these 
final yields.  This so-called ``$\nu$-process''\cite{woosley90} can produce rare isotopes that 
are potentially detectable in both stellar and meteoritic data.
For example, the raio of 92Nb to 93Nb has been measured using solar
system meteorites\cite{hayakawa13}

Because the nucleosynthetic yields are typically integrated results,
what we learn from them depends heavily on our theoretical understanding 
of both the supernova engine and nuclear physics.  We must use all of 
our diagnostics to help constrain the physics behind supernovae before 
we can truly take advantage of the integrated nucleosynthetic yields.  
But with a basic theoretical understanding calibrated by our other 
diagnostics, we can use these yields to place strong constraints 
on nuclear burning rates and the origin of the heavy elements.

\section{Tying multiple constraints together, Post-explosion nucleosynthesis}
\label{sec:postexp}

Understanding the r-process is an example of how multiple diagnostics
can be used together in concert with theory to study a physical
process.  The origin of the r-process elements remains one of the
great unsolved problems of nuclear astrophysics with many production
sites with neutron star mergers\cite{levinson93} and core-collapse
supernovae\cite{woosley92} being dominant sources.  Although it is
possible that the merger of two neutron stars may be the dominant
source of the r-process, it is likely that core-collapse supernovae
play some role in the production of these heavy elements.  After the
launch of the explosion, neutrinos leak out of the neutron star,
driving a neutron-rich wind.  Neutron capture in this wind can build
up extremely heavy elements, and the neutrino-driven wind mechanism
has long been considered a prime site for r-process nucleosynthesis.
Post-explosion nucleosynthesis focuses on the nucleosynthetic yields
from this cooling neutron star.

The basic neutrino-drive wind r-process model argues that the high
neutrino flux of the cooling neutron star ejects the outer layer of
the neutron star (the ejecta mass need only be $10^{-6}$\,M$_\odot$ to
explain r-process yields)\cite{woosley92,meyer92,
  howard93,takahashi94,woosley94,qian96}.  The anti-electron neutrinos
are typically higher-energy than the electron neutrinos and hence have
a higher cross-section with the wind matter.  This preserves the
neutron-rich nature of this ejecta and rapid neutron capture produces
the r-process.  The high entropies, fast expansion, and neutron
richness of this wind ejecta may provide the right conditions for
making r-process nuclei\cite{qian96}.  However, current calculations
suggest that the basic model for neutrino-driven wind r-process only
works in extreme conditions (e.g., $>$2\,M$_\odot$ neutron
stars\cite{argast04,suzuki05}, for a review, see\cite{thompson01}).

A number of solutions have been proposed, including magnetic fields
that can alter the wind trajectories, providing time for sufficient
neutron captures and decays to produce the
r-process\cite{suzuki05,metzger07}.  But most of the detailed physics
work has focused on the role nuclear physics might play in ``fixing''
the neutrino-driven wind site for r-process.  Unfortunately,
self-consistent studies including neutrino capture reactions into a
reaction network code have shown that that the interplay between
neutrino and nuclear reactions, if anything, makes it more difficult
to produce the r-process\cite{fuller95,mclaughlin96,meyer98}.
Neutrino oscillations (including active-sterile neutrino conversion)
have been proposed as a nuclear-physics solution fixing the
neutrino-driven r-process scenario\cite{fuller97,fetter03}.  The
uncertainties in the nuclear mass models still can change this picture
dramatically\cite{xu13}.  With the development of the Facility for Rare
Isotope Beams (FRIB), this physics uncertainty may be removed,
strengthening the studies of neutrino physics.

\begin{figure}
\includegraphics[width=8.0cm]{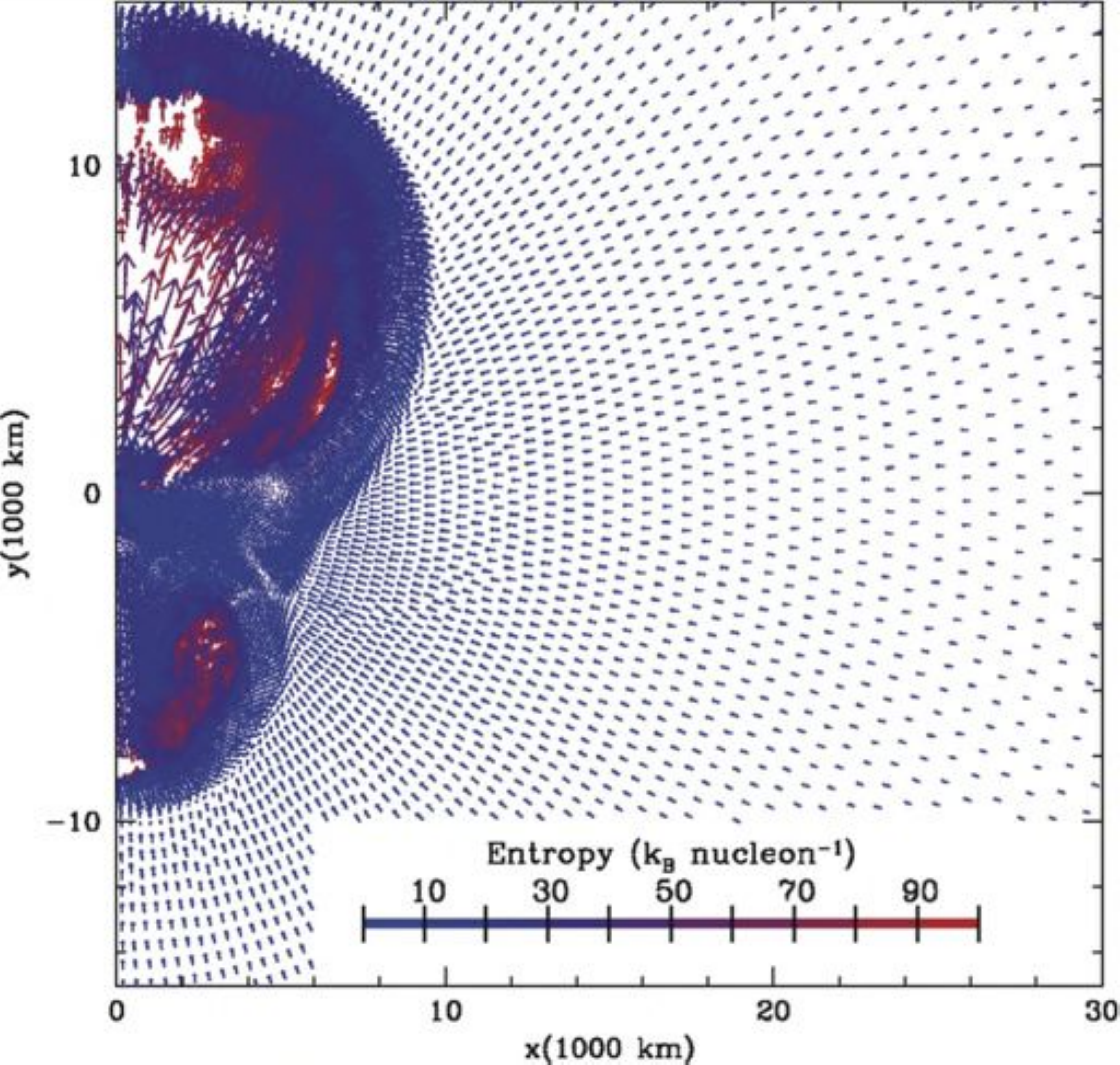}
\caption{A snapshot in time of a 2-dimensional fallback accretion
  model (the neutron star is at the 0,0 point).  The arrows denote the
  direction of the velocity and the outflow can be seen along the y
  axis.  Colors denote entropy and the entropy of the ejecta can be
  very high entropy\cite{fryer06c}.  \label{fig:postfb}}
\end{figure}

An alternate solution invokes the fallback of material after the
launch of the explosion.  Fallback material crashes back down onto the
proto-neutron star, providing a new source of neutrinos and heat.
Some (up to 25\%) of this material can be ejected\cite{fryer06c}.
Figure~\ref{fig:postfb} is a snapshot in time during this fallback,
showing the outflowing ejecta.  Much of the re-ejected material is not
neutron rich and typically produces iron-peak elements.  This ejecta
provides a second source of $^{56}$Ni to help power supernova
light-curves.  Since most of the fallback occurs in the first 100\,s,
this new ejecta becomes part of the sueprnova explosion when the shock
breaks out of the star.  The energy from this fallback ejecta,
espeically late-time fallback, has been used to explain a variety of
peculiar supernovae\cite{dexter13}.

The yield of this ejecta is highly sensitive to the amount of free
neutrons in the system.  A 1\% variation in the electron fraction can
make the difference between a clear r-process signature (with an
electron fraction of 0.495) and a strong iron peak yield (with an
electron fraction of 0.505).  The evolution of an r-process yield in
this fallback scenario is very different than typical r-process
production sites.  With the low neutron fraction of this scenario, the
abundances stay near the the valley of stability and proton capture is
able to drive the element mass beyond waiting points.  Only at late
times do the protons freeze out, allowing the neutrons to develop
a typical r-process isotope distribution.  This nucleosynthetic path
was termed ``rpn-process'' (rapid proton and neutron capture) in the
first fallback study\cite{fryer06c} and it is a variant of the
i-process (intermediate process) heavy element nucleosynthesis.
Because of the sensitivity to the exact number of neutron and proton
fractions, this site is sensitive to the hydrodynamics, neutrino
processes and nuclear reaction rates and much more must be studied to
understand it better.

To test this model, we can compare our r-process signature to the
observed signature in stars or meteorites (see Sec.\ref{sec:nucobs}).
If we trust our fallback model implicitly, we can use these yields to
constrain the nuclear physics governing these yields.  Independent
validation of our fallback models is needed to provide that trust.  Of
course, the neutrino signal will measure this late-time fallback.  But
we can also utilize the data on compact remnants and supernova light
curves to validate our fallback model.  Let's review this validation 
data in more detail.

\begin{figure}
\includegraphics[width=8.0cm]{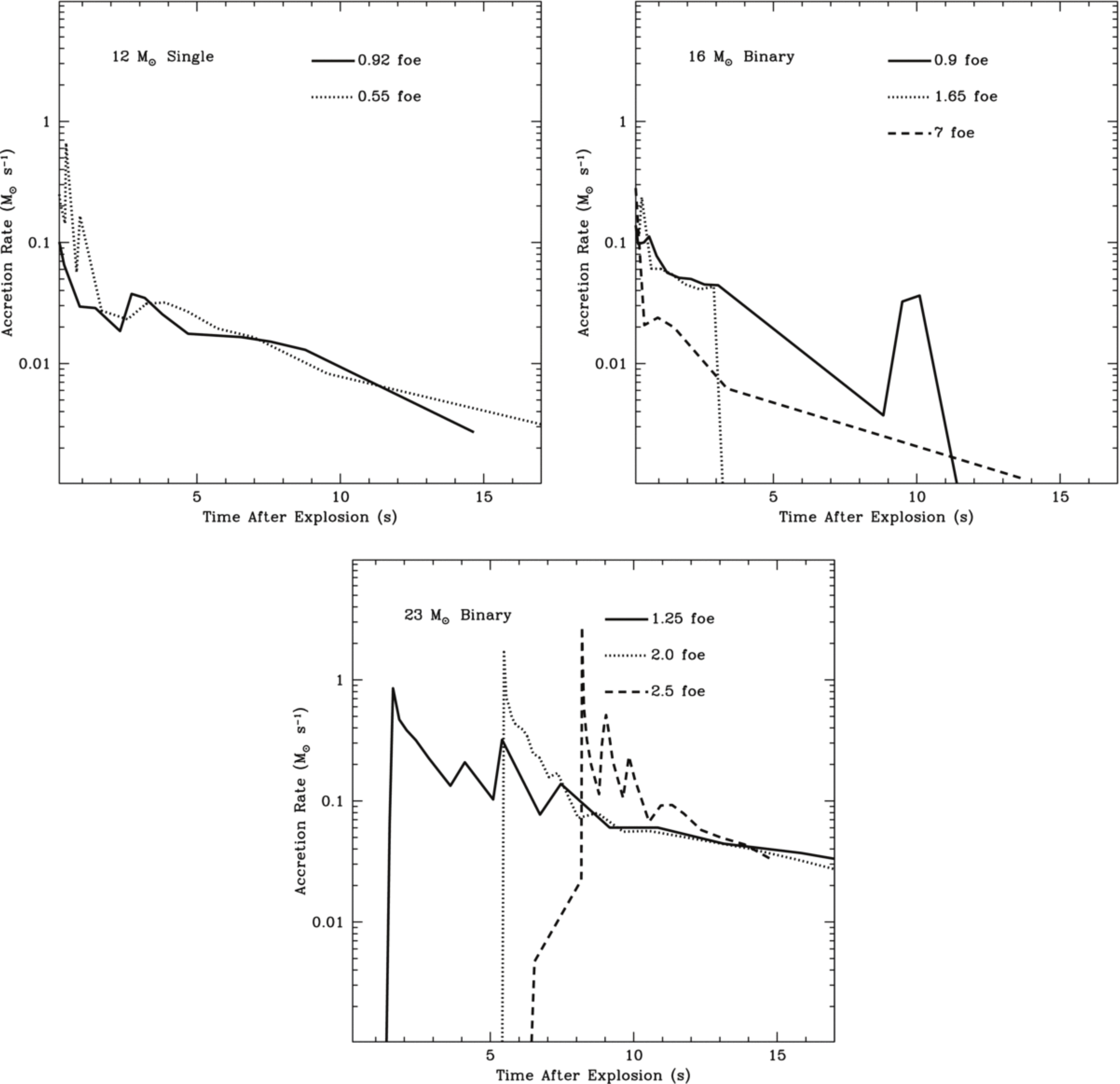}
\caption{The fallback accretion rate for 3 different progenitors each
  with different explosion energies\cite{fryer09}.  The accretion
  stars nearly immediately after the launch of the explosion.  Indeed,
  in many core-collapse engine models, this fallback can be
  seen while the shock moves outward.\label{fig:postfbrate}}
\end{figure}

Because current explosive engine calculations are still sensitive to
broad range of details (see Hix review), we can not hope to cover the
possible range of explosion features to conduct a complete study of
fallback.  Instead, most calculations implement artificially the
explosion energy.  Unfortunately, how this energy is injected can
drastically alter the amount and timing of the fallback.  Models
driven by pistons tend to produce late time fallback whereas the
arguably more physically-relevant energy driven explosions produce
fallback almost immediately\cite{young07}.
Figure~\ref{fig:postfbrate} shows the fallback rates estimated for 3
different progenitors and a range of explosion energies\cite{fryer09}.  For these
calculations, the fallback occurs immediately.  It is this fallback
that produces the r-process yields, but only if there is considerable
fallback.  We can use studies of compact remnant masses to get a
handle on the amount of fallback in typical supernovae.  The amount of
systems that form remnant masses above 1.4-1.5\,M$_\odot$ provides us
with a rough estimate of how many systems are likely to have more than
a few tenths of a solar mass of fallback.  Comparing these results to
our fallback models can determine the rate at which supernovae can 
produce r-process from fallback.

We can also try to use supernova light-curves to constrain fallback.  
With a supernova light-curve, we can estimate the explosion energy and 
the role the fallback accretion plays in the light-curve.  If we can 
see evidence of the fallback in the light-curve, we will be able to 
place constraints on the nature of the fallback.  Unfortunately, although 
fallback has been argued to explain some peculiar supernovae\cite{dexter13}, 
the fact that alternation explanations exist for these supernovae make it 
difficult to use this as a constraint.  However, if we can produced an 
unbiased fraction of weak supernovae (with little $^{56}$Ni yield), the 
fraction can be used to better understand fallback.  Coupled with the 
remnant mass distribution, these observations will be able to distinguish 
between early and late-time fallback mechanisms.

With a neutrino detection following the fallback, we will be able 
to strongly constrain the fallback and for nearby supernovae where all of 
our diagnostics are available, we should be able to make strong predictions 
on the total amount of fallback and its timing.

\section{summary}

Supernovae are indeed Nature's laboratories for matter in extreme
conditions.  Traditional studies of this extreme physics have focused
on the neutrino and gravitational wave diagnostics of supernovae.  But
nearly all the data we have on supernovae used different diagnostics.
In this review, we have discussed the range of diagnostics: neutrinos,
gravitational waves, progenitor studies, the supernova outbursts (a
wide range of photon energies), compact remnant mass distributions,
ejecta remnant observations, and integrated nucleosynthetic yield 
constraints from low-metallicity stars and meteoritics.

All of these diagnostics have strengths and weaknesses, but taken
together, we are at the stage of truly constraining nuclear and
particle physics with our supernova ``experiment''.  Some of these
diagnostics require considerable theory work to interpret and any
study of supernovae as physics laboratories will require a strong
theoretical understanding.  The wealth of data will be able to both
help us understand the uncertainties in this theoretical understanding
as well as provide constraints on our errors in studying nuclear
physics.

Supernova studies are at a critical point where new surveys and 
new observational experiments will soon drastically increase the 
amount of data.  Along with advances in our theoretical understanding, 
we are on the cusp of a revolution in this science and it is likely 
that the next decade will see large breakthroughs in this science.

\begin{acknowledgments}

The work of C.L.F. and W.E. was done under the auspices of the
National Nuclear Security Administration of the U.S. Department of
Energy at Los Alamos National Laboratory under Contract
DE-AC52-06NA25396.

\end{acknowledgments}


\end{document}